\documentclass[%
 reprint,
superscriptaddress,
 amsmath,amssymb,
aps,
pre,
floatfix,
]{revtex4-1}

\usepackage[utf8]{inputenc}
\usepackage{graphicx}
\usepackage{dcolumn}
\usepackage{bm}
\usepackage{bbm}
\usepackage{subfig}
\usepackage{float}

\begin{document}

\title{Systemic Performance Measures from Distributional Zeta-Function}

\author{C. D. Rodr\'iguez-Camargo}
 \email{cdrodriguezc@unal.edu.co}
\affiliation{%
Centro de Estudios Industriales y Log\'isticos para la productividad (CEIL, MD)\\
Programa de Ingenier\'ia Industrial\\
Corporaci\'on Universitaria Minuto de Dios, Bogot\'a AA 111021, Colombia
}%
\affiliation{%
 Programa de Investigaci\'on sobre Adquisici\'on y An\'alisis de Se\~nales (PAAS-UN)\\ Universidad Nacional de Colombia, Bogot\'a AA 055051, Colombia
 }%
\author{A. F. Urquijo-Rodr\'iguez}
\email{afurquijor@unal.edu.co}
\affiliation{%
Grupo de Superconductividad y nanotecnolog\'ia \\
Departamento de F\'isica\\
Facultad de Ciencias\\
Universidad Nacional de Colombia, Bogot\'a AA 055051, Colombia
}%

\author{E. A. Mojica-Nava}
\email{eamojican@unal.edu.co}
\affiliation{%
Departamento de Ingenier\'ia El\'ectrica y Electr\'onica \\ 
Facultad de Ingenier\'ia \\ 
Universidad Nacional de Colombia, Bogot\'a AA 055051, Colombia
}%
\affiliation{%
 Programa de Investigaci\'on sobre Adquisici\'on y An\'alisis de Se\~nales (PAAS-UN)\\ Universidad Nacional de Colombia, Bogot\'a AA 055051, Colombia
 }%

\begin{abstract}
We propose the use of the Distributional Zeta-Function (DZF) for constructing a new set of Systemic Performance Measures (SPM). SPM have been proposed to investigate network synthesis problems such as the growing of linear consensus networks. The adoption of the DZF has shown interesting physical consequences that in the usual replica method are still unclarified, i.e., the connection between the spontaneous symmetry breaking mechanism and the structure of the replica space in the disordered model. We relate topology of the network and the partition function present in the DZF by using the spectral and the Hamiltonian structure of the system. The studied objects are the generalized partition funcion, the DZF, the Expected value of the replica partition function, and the quenched free energy of a field network. We show that with these objects we need few operations to increase the percentage of performance enhancement of a network. Furthermore, we evalue the location of the optimal added links for each new SPM and calculate the performance improvement of the new network for each new SPM via the spectral zeta function, $\mathcal{H}_{2}$-norm, and the communicability between nodes. We present the advantages of this new set of SPM in the network synthesis and we propose other methods for using the DZF to explore some issues such as disorder, critical phenomena, finite-temperature, and finite-size effects on networks. Relevance of the results are discussed. \\
\end{abstract}

\maketitle

\section{Introduction}
A topic of special interest in network science is the improvement of robustness and global performance in order to respond adequately to external disturbances~\cite{wu1, abbas1, siami1, wang1, ye1, duan1, liang1, pizzuti1, wang2, shang1, wang3, lu1, pizzuti2}. These issues are crucial for the sustainability of large scale dynamical networks from engineering to biological infrastructures~\cite{girvan1, mason1, zhang1}. One of the essential problems in this area is to study the effect of the uncertain exogenous inputs over remoteness of perturbed trajectories with respect to its working equilibrium point. The primary challenge, in these kind of problems, is to introduce meaningful and viable performance and robustness measures. These objects must capturate essential characteristics of the network. An accurate measure should be able to encapsulate steady-state, transient, microscopic, and macroscopic features of the perturbed large-scale dynamical network.

The performance analysis of linear consensus networks exposed to external stochastic disturbances has been studied by different objects. For example the $\mathcal{H}_{2}$-norm of the network has been employed as a scalar performance measure that captures the concep of coherence~\cite{bamieh1}. An important result about the structure of these objects shows that $\mathcal{H}_{2}$-norm is a function of the eigenvalues of the Laplacian matrix~\cite{bamieh1, young1, siami2} under certain conditions (i.e., If the Laplacian matrix of the coupling graph of the network is normal). Besides the $\mathcal{H}_{2}$-norm, there are other functions presented as performance measures in~\cite{bamieh1}, \cite{zelazo1, ofa1}. In~\cite{siami3}, it is shown that these objects acting as performance measures are Schur-convex functions in regard to the Laplacian eigenvalues. These performance measures, that are defined from entropy, spectral, and some system norms functions, manifest several helpful functional properties that permit their usage in network synthesis problems~\cite{siami4}.

Recently, a kind of spectral function of Laplacian eigenvalues called Systemic Performance Measures (SPM) has been proposed to investigate network synthesis problems, such as the growing of linear consensus networks~\cite{siami5}. Numerous and widely used performance measures belong to this class, for example, spectral zeta function, Gamma entropy, expected transient output covariance, system Hankel norm, convergence rate to consensus state, logarithm of uncertainty volume of the output, Hardy-Schatten system norm, and many more. All these objects are monotone, convex, and orthogonally invariant~\cite{siami5}.

In network analysis and especially for those methods based on statistical physics analogies, most adopted approaches to study complex structures are based on concepts from spectral graph theory~\cite{minello0}. It is known that from the analogies based on statistical mechanics~\cite{barabasi1, bianconi0, estrada1, park1, javarone1, bianconi1, ostilli1, bianconi2}, thermodynamics~\cite{escolano1, ye11, ye2, minello1, minello2, wang4, wang5}, as well quantum information~\cite{braun1, paserini1, anand1, minello4},  we can extract a set of spectral functions of Laplacian eigenvalues that are related with ensemble and thermodynamic quantities of interest. When a network is described by a partition function, thermodynamic quantities, such as energy, temperature and entropy can be straightforwardly derived from it. There are various approaches to obtain the partition function of a network~\cite{gabrielli1}. For instance, using the heat bath analogy, where the energy states are related to the eigenvalues of a matrix representation of network structure, particles, which are in thermal equilibrium with the heat bath, begin to populate these energy states. Within this thermalization process, the energy states can be described by Maxwell–Boltzmann~\cite{wang5, metz1}, Bose–Einstein~\cite{wang5, bianconi0}, and Fermi–Dirac~\cite{wang5, bianconi3, shen1, moura1, murphy1} occupation statistics. On the other hand, in~\cite{ye11} the authors show that the partition function can be computed from the matrix characteristic polynomial. In other works, such as in~\cite{anand1, minello4} the reduced Laplacian graph matrix $\tilde{L}$ is associated with the density matrix $\hat{\rho}$, i.e., $\hat{\rho}\sim \tilde{L}$. However, the most used approach is establishing an identity relationship between the adjacency or Laplacian matrix, $A$ and $L$, respectively, with the Hamiltonian operator $\hat{H}$, i.e., if $\hat{H}=-\Delta+U(r,t)$ then the operators might be $L=-\Delta$ and $\hat{H}=-A$, or simply $\hat{H}=L$, being $\Delta$ the generalized Laplacian operator.

Using the aforementioned approach, we construct a set of Laplacian spectral functions from the formalism of Distributional Zeta Function (DZF) that may serve us as SPM.

The DZF is a recent alternative method to average the disorder-dependent free energy in statistical field theory~\cite{svaiter1}. Within this approach, the dominant contribution to the average free energy is expressed as a series of the integer moments of the partition function of the model. The adoption of the indicated alternative has shown interesting physical consequences that have been unnoticed by the usual techniques such as the cavity~\cite{mezard1, mezard2} and replica methods~\cite{edw1}, where the concept of replica symmetry breaking was introduced by Parisi in virtue of prevent unphysical results~\cite{parisi1, parisi2, parisi3, parisi4}. In the framework of DZF, it is proved the connection that exists between the spontaneous symmetry breaking mechanism and the replica symmetry ansatz in a disordered scalar model~\cite{svaiter2}. In~\cite{svaiter2}, the authors show that since all replica partition functions are making a contribution to average free energy, the only alternative in each replica partition function is the replica symmetric ansatz, where according to DZF method the system has the possibility to develop a spontaneous symmetry breaking. We use this fact to generate objects where each subsystem of a complete graph is contributing to improve the network performance measure. The DZF method has been used successfully to study the Landau-Ginzburg approach in replica field theory~\cite{svaiter2}, the disordered $\lambda \varphi ^{4}+\rho \varphi ^{6}$ Landau-Ginzburg model~\cite{svaiter3}, disordered Bose-Einstein condensate in hard walls trap~\cite{svaiter4}, multiplicative noise in Euclidean Schwarzschild manifold~\cite{svaiter5}, and more recently for polymers in random media~\cite{svaiter6}. Furthermore, it has been mentioned its potential for entanglement networks in random media~\cite{svaiter7}.

In this paper, using auxiliary Euclidean fields over a disorder-induced interaction network, we construct a new set of SPM from DZF approach and we examine its advantages respect to other performance measures for searching meaninful and viable performance and robustness measures, and then for network synthesis problems. We study a $d$-dimensional Euclidean field theory where the replica fields are interacting via an external disordered field $h(x)$. The objects subjected to study are the following: i) The generalized field network partition funcion; ii) The DZF of a field network; iii) The expected value of the replica field network partition function; and iv) The quenched free energy of a field network. We take advantage from the disordered-induced interaction, the functional form of DZF, the series representation of quenched free energy, and the spontaneous symmetry breaking mechanism and its physical implications to generate a set of parameters that maximize the improvement action of each new SPM. We show that with these objects we can obtain a performance improvement with a few operations over the network. Furthermore, we trace a path for the study of random media, finite size and temperature effects in this framework, beside a way to explore new issues in network science by statistical field theory.

This paper is organized as follows: In Sec.\ref{preliminaries}, we present the definitions and backgrounds neccessary to tackle our problem. We revisit the SPM and DZF structure and define the objects to be extended to networks. In Sec. \ref{dzfpm}, we show the detailed construction of each aforementioned element of our new set of SPM and its conditions to be a SPM for the synthesis network problem. In Sec. \ref{numericalresults}, we discuss the numerical results where we show that this new set is exposing advantages respect to the known SPM such as spectral zeta function and spectral entropy. We evalue the percentage of performance enhancement for a given quantity of links added to the original network. Also, we evalue the location of the optimal added links for each new SPM and calculate the performance improvement of the new graph for each new SPM via the spectral zeta function~\cite{siami3}, $\mathcal{H}_{2}$-norm~\cite{bamieh1}, and the communicability between nodes~\cite{estrada2}. The concluding remarks are given in Sec. \ref{conclusions}. Theorems of interest are consigned in Appendix.

\section{Preliminaries and Definitions}
\label{preliminaries}
In this section, we review the background framework of our SPM construction from DZF approach. We begin by defining graphs and SPM, then explain the DZF method, and finally show a way to relate them.
\subsection{Graphs}
A graph is a 3-tuple $G=[V,E,w]$ consisting of a set of \emph{nodes} or \emph{vertex} $k\in V$, \emph{links} or \emph{edges} $(i,j)\in E \subseteq V \times V$ and a function $w:E\rightarrow \mathbb{R}^{+}$ which assigns a weight to each edge. A network is a 3-tuple $G(t)=[V(t),E(t),w(t);J]$. The dynamic behaviors parametrized by the temporal dimension $t$ are called microrules that are represented algorithmically by $J$. A Network is a dynamical graph.

A graph can be represented by its (weighted) adjacency matrix $W$ as $W_{ij}=w(i,j)$, being $w(i,j)$ the weight of the link between the nodes $v_{i}$ and $v_{j}$. For a unweighted network we adopt the usual notation $A$ and $a_{ij}$ for its elements. The (weighted) degree matrix is defined by $D=\text{diag}(d_{1}\cdots d_{k})$ where the (weighted) degree $d_{i}$ of a node $v_{i}$ is given by
\begin{equation}
d_{i}=\sum _{n=1}^{k}w(i,n) ,
\end{equation}
being $k$ the number of nodes in a graph.

The Laplacian matrix is defined by
\begin{equation}
L=D-W.
\label{laplacianmatrix000}
\end{equation}

The normalized Laplacian matrix $\tilde{L}$ is defined as
\begin{equation}
\tilde{L}=D^{-1/2}LD^{-1/2}
\label{laplacianmatrix111}
\end{equation}

The elementwise expression of $\tilde{L}$ is
\begin{equation}
\tilde{l}_{ij} =\begin{cases}
1\quad \text{if}\quad i=j \qquad \text{and}\quad d_{j}\neq 0,\\
-\frac{1}{\sqrt{d_{i}d_{j}}} \quad \text{if}\quad i\neq j \quad \text{and}\quad (i,j)\in E,\\
0 \qquad \text{otherwise}.
\end{cases}
\label{tildee}
\end{equation}

We denote the set of all Laplacian matrices that represent $k$-nodes connected weighted graphs by $\mathcal{L}_{k}$. Since $G$ is both connected and undirected, the Laplacian matrix $L$ has $k-1$ strictly positive eigenvalues and one zero eigenvalue. Taking $0=\lambda _{1}<\lambda _{2}\leq \cdots \leq \lambda _{k}$ as the eigenvalues of Laplacian matrix $L$, the operator $\Lambda ^{(k)}:\mathbb{S}^{k}_{+}\rightarrow \mathbb{R}^{k-1}_{+}$ is defined by $\Lambda ^{(k)}(L)=[\lambda _{2}\cdots \lambda _{k}]^{T}$. The Moore-Penrose pseudoinverse of $L$ is written as $L^{\dagger}=[l^{\dagger}_{ij}]$, which is a symmetric, doubly centered, square, and positive-semidefinite matrix. The eigenvalues of (\ref{laplacianmatrix111}) are denoted by $\tilde{\lambda}_{i}$. For a given link $e=\{ i,j \}$, $r_{e}(L)$ indicates the effective resistance bewteen nodes $i$ and $j$ in a graph represented by a Laplacian matrix $L$, where its value can be computed by
\begin{equation}
r_{e}(L)=l_{ii}^{\dagger}+l_{jj}^{\dagger}-2l_{ij}^{\dagger} .
\end{equation}

\emph{Definition 2.1:} The derivative of a scalar function $\rho (.)$, with respect to the matrix $X\in \mathbb{R}^{k\times k}$, is defined by
\begin{equation}
\nabla _{X} \rho (X) := \begin{pmatrix}
\frac{\partial \rho}{\partial x_{11}} && \frac{\partial \rho}{\partial x_{12}} && \cdots && \frac{\partial \rho}{\partial x_{1k}} \\
\frac{\partial \rho}{\partial x_{21}} && \frac{\partial \rho}{\partial x_{22}} && \cdots && \frac{\partial \rho}{\partial x_{2k}} \\
\vdots && \vdots && \ddots && \vdots \\
\frac{\partial \rho}{\partial x_{k1}} && \frac{\partial \rho}{\partial x_{k2}} && \cdots && \frac{\partial \rho}{\partial x_{kk}} \\
\end{pmatrix} ,
\end{equation}
where $X=[x_{ij}]$. The directional derivative of function $\rho (X)$ in the direction of matrix $Y$ is given by $\nabla _{X,Y}\rho (X)=\langle \nabla \rho (X),Y\rangle = \text{Tr}(\nabla \rho (X) Y)$.

\emph{Definition 2.2.:} For every $x \in \mathbb{R}^{n}_{+}$, we denote $x^{\downarrow}$ as a vector whose elements are a permuted version of elements of $x$ in descending order. We say that $x$ majorizes $y$, which is represented by $x \trianglerighteq y$, if and only if $\mathbbm{1}^{T}x=\mathbbm{1}^{T}y$ and $\sum _{i=1}^{k}x^{\downarrow}_{i}\geq \sum _{i=1}^{k}y^{\downarrow}_{i}$, $\forall k=1,...,n-1$~\cite{marshall1}.

\emph{Definition 2.3:} The real-valued function $F:\mathbb{R}^{n}_{+}\rightarrow \mathbb{R}$ is called Schur convex if $F(x)\geq F(y)$ for every two vectors $x$ and $y$ with property $x \trianglerighteq y$~\cite{marshall1}.

\emph{Definition 2.4:} The usual Kronecker product is defined by  
\begin{equation}
A \otimes B = \begin{pmatrix}
a_{11}B && \cdots && a_{1n}B \\
\vdots && \ddots && \vdots \\
a_{m1}B && \cdots && a_{mn}B
\end{pmatrix},
\end{equation} 
being $A\in \mathbb{R}^{m\times n}$, $B\in \mathbb{R}^{p\times q}$ and $A \otimes B \in \mathbb{R}^{pm \times qn}$.

\emph{Definition 2.5:} Assume that the fractal dimension $D$ is written as $D=n\, (\text{integer})+d\,(\text{decimal})$. Then we define the fractal dimensional matrix as a usual matrix where we add a special decimal row (column)~\cite{chang11}. The $D$-dimensional square matrix is thus
\begin{equation}
\begin{pmatrix}
a_{11} && \cdots && a_{1n} && a_{1D}\\
\vdots && \ddots && \vdots && \vdots \\
a_{n1} && \cdots && a_{nn} && a_{nD}\\
a_{D1} && \cdots && a_{Dn} && a_{DD}
\end{pmatrix}
\label{fractal111}
\end{equation} 
in which the final row and column should be understood as a special decimal dimension $d$. So all of linear algebra can be applied by the same way, but difference is only the final row and column.
\subsection{Systemic Performance Measures}

A SPM is defined as a real valued operator defined over the set of all linear consensus networks determined by the following expressions
\begin{eqnarray}
\dot{x}(t)&=&-Lx(t)+\eta (t), 
\nonumber \\
y(t)&=&M_{k}x(t),
\end{eqnarray}
where $x=[x_{1},...,x_{k}]^{T}$ is the vector state variable, $y=[y_{1},...,y_{k}]^{T}$ is the output, $\eta = [\eta _{1},...,\eta _{k}]$ a exogenous noise input, $L$ is a graph Laplacian matrix defined by (\ref{laplacianmatrix000}), and $M_{k}$ the output matrix given by $M_{k}:=I_{k}-\frac{1}{k}J_{k}$, that quantifies the quality of noise propagation in these networks, being $I_{k}$ and $J_{k}$, the $k\times k$ identity matrix and $k \times k$ matrix of all ones, respectively. 

An operator $\rho :\mathcal{L}_{k}\rightarrow \mathbb{R}$ is called a SPM if it satisfies the following properties $\forall L \in \mathcal{L}_{k}$: i) \emph{Monotonicity:} If $L_{2}\preceq L_{1}$, then $\rho (L_{1})\leq \rho (L_{2})$; ii) \emph{Convexity:} $\forall \alpha$ with $\quad 0\leq \alpha \leq 1$, we have $\rho (\alpha L_{1}+(1-\alpha )L_{2})\leq \alpha \rho (L_{1})+(1-\alpha )\rho (L_{2})$; iii) \emph{Orthogonal invariance:} For all orthogonal matrices $U\in \mathbb{R}^{k\times k}$, we have $\rho (L)=\rho (ULU^{T})$ (see definition 4 in~\cite{siami5}).

Furthermore, this operator $\rho :\mathcal{L}_{k}\rightarrow \mathbb{R}$, with these properties, is indeed a Schur-convex function of Laplacian eigenvalues and it can be represented by a a Schur-convex spectral function $\Phi :\mathbb{R}^{k-1}\rightarrow \mathbb{R}$ such that $\rho (L)=\Phi (\lambda _{2},...,\lambda _{k})$ (see Theorem 1 in \cite{siami5}). See the Appendix (\ref{appendix2}) to explore subsequent results which we will use to determine the conditions that a selected spectral function must accomplish to be a SPM.

Some important examples of spectral systemic performance measures and its matrix operator and spectral representation are shown:
\begin{itemize}
\item Spectral zeta function
\begin{eqnarray}
\zeta _{q}(L)&=&(\text{tr}((L^{\dagger})^q))^{1/q} 
\nonumber \\
&\mapsto & \left( \sum _{i=2}^{k}\lambda _{i}^{-q}\right) ^{1/q}
\label{spectralzetafunction1}
\end{eqnarray}
\item Gamma entropy
\begin{eqnarray}
I_{\gamma}(L)&=&\gamma ^{2}\text{tr}\left( L-(L^{2}-\gamma ^{2}M_{n})^{1/2}\right) 
\nonumber \\
&\mapsto &\gamma ^{2}\sum _{i=2}^{k}\lambda _{i}-(\lambda _{i} ^{2}-\gamma ^{-2})^{1/2}
\label{gammaentropy1}
\end{eqnarray}
\item Expected transient output covariance
\begin{eqnarray}
\tau _{t}(L)&=&\frac{1}{2}\text{tr}(L^{\dagger}(I-e^{-Lt}))
\nonumber \\
& \mapsto & \frac{1}{2}\sum _{i=2}^{k}\lambda _{i}^{-1}(1-e^{-\lambda _{i} t})
\label{expectedtransientoutput}
\end{eqnarray}
\item Uncertainty volume of the output 
\begin{eqnarray}
v(L)&=&(1-k)\text{log}2-\text{tr}\left( \text{log}\left( L+\frac{1}{k}J_{k}\right) \right) 
\nonumber \\
&\mapsto &(1-k)\text{log}2-\sum _{i=2}^{k}\text{log}\lambda _{i}
\label{uncertaintyvolumeoftheoutput}
\end{eqnarray}
\end{itemize}
In~\cite{siami5}, beside show that the above functions are SPM, the authors use them for growing linear consensus networks and improve its properties. In this paper we use these algorithms to explore the sensivity of location of optimal links for each proposed SPM and its induced growing pattern. See Appendix (\ref{subappend111}) for further details.
\subsection{The distributional zeta-function in disordered field theory}
In this subsection we revisit how to obtain a replica field theory from an Euclidean scalar field theory in the presence of a disorder field. From functional integral formulation of field theory, we have two kinds of random variables. Primarily, we have the Euclidean fields. These fields are describing generalized Euclidean processes with zero mean and a covariance defined in terms of gradients. Apart from that, the other random variables are the disorder fields, characterized by the absence of any differential operator. Here we shall follow the description given in~\cite{svaiter1, svaiter2}.

The functional integral of the Euclidean scalar $\lambda \varphi ^{4}$ model in the presence of a disorder field $h(x)$ is defined by
\begin{equation}
Z(h)=\int d[\varphi]\text{exp}\left( -S+\int d^{d}xh(x)\varphi (x)\right),
\label{funct1}
\end{equation}
being $S=S_{0}+S_{I}$ the action that usually depicts a massive scalar field, where the contribution $S_{0}$ is given by
\begin{equation}
S_{0}(\varphi)=\int d^{d}x\left( \frac{1}{2}(\partial \varphi)^{2}+\frac{1}{2}m_{0}^{2}\varphi ^{2}(x)\right)
\end{equation}
and the contribution $S_{I}$ given by
\begin{equation}
S_{I}(\varphi)=\int d^{d}x \frac{g_{0}}{4!}\varphi ^{4}(x) .
\end{equation}

The term $S_{0}$ is the free-field action, while $S_{I}$ is a non-Gaussian contribution that accounts for the interacting component. In the expression (\ref{funct1}), $d[\varphi]$ is a functional measure given by 
\begin{equation}
d[\varphi]=\prod _{x}d\varphi (x) .
\end{equation}

The terms $g_{0}$ and $m_{0}^{2}$ are the bare coupling constant and the bare mass square of the model, respectively. Finally, $h(x)$ is a quenched random field, with probability distribution $d\rho [h]=d[h]P(h)$, where $P(h)$ is
\begin{equation}
P(h)=p_{0}\text{exp}\left( -\frac{1}{2\sigma}\int d^{d}x(h(x))^{2}\right) .
\label{prob1}
\end{equation}

The constant $\sigma$ is a small positive parameter associated with the disorder, while $p_{0}$ is a normalization factor. This is a delta-correlated field, i.e, $\mathbb{E}(h(x)h(y))=\sigma \delta ^{d}(x-y)$.

In these kind of scenarios, it is necessary to eliminate the disorder field. For a specified probability distribution $P(h)$, we may average the disordered functional integral $Z(h)$, and after apply the logarithm, yielding the definition of the annealed free energy. There is another free energy, that is called the quenched free energy and it is defined by
\begin{equation}
F_{q}=-\int d[h]P(h)\ln Z(h) ,
\label{lq}
\end{equation}
being $d[h]=\prod _{x}dh(x)$ a formal Lebesgue measure. 

For computing (\ref{lq}), usually it is employed the replica method. The main point in this method is to compute integer moments of partition function $\mathbb{E}[Z^{k}]$ and use such information to calculate $\mathbb{E}(\ln Z)$. In this method, firstly, we construct the $k$-th power of the partition function $Z^{k}=Z\times Z \times \cdots \times Z$. We interpret that product as a new system formed by $k$ statistically independent copies of the original system. Next, the expected value of the partition function's $k$-th power $\mathbb{E}[Z^{k}(h)]$ is evaluated by integrating over the disorder field on the new model (collection of replicas). Notice that in $Z^{k}$, integration over disorder field yields a system defined by $k$ replicas which are no more statistically independent. Finally, the average free energy is computed using the following identity
\begin{equation}
\mathbb{E}[\ln Z(h)]= \lim _{k\rightarrow 0}\frac{\partial}{\partial k}\mathbb{E}[Z^{k}(h)] .
\end{equation}

The average value in the presence of the quenched disorder is then obtained in the limit of a zero-component field theory, taking the limit $k\rightarrow 0$. 

The alternative approach to calculate (\ref{lq}) is presented in~\cite{svaiter1, svaiter2} and it is called the distributional zeta funcion. An interesting issue of this method is that it is possible to find an analytic expression for the free energy and all the replicas are included on it. Their analysis starts from the definition of a generalized zeta function 
\begin{equation*}
\zeta _{\mu , f}(s)=\int _{X}f^{-s}(x)d\mu (x),
\end{equation*}
where $(X,\mathcal{A}, \mu)$ is a measure space and $f:X \rightarrow (0,\infty)$ is measurable. Therefore, for example if $f(x)=x$, $X=\mathbb{N}$ and $\mu$ being the counting measure, the Riemann zeta function is obtained~\cite{riemann1, ing1}. If $\mu$ counts only the prime numbers, we have the prime zeta function~\cite{ulandau1, froberg1}. If $X=\mathbb{R}$ and $\mu$ counts the eigenvalues of an elliptic operator, the spectral zeta function is obtained~\cite{blau1, haw1, hajli1}. The authors in~\cite{svaiter1, svaiter2} extend this formalism for the case $f(h)=Z(h)$ and $d\mu (h)=d[h]P(h)$ leading the definition of the DZF as
\begin{equation}
\Phi (s)=\int d[h]P(h)\frac{1}{Z^{s}(h)} .
\end{equation}

Following the common steps for the spectral zeta function, the average free energy $F_{q}$ can be written as 
\begin{equation}
F_{q}=\left. \frac{d}{ds}\Phi (s)\right| _{s=0^{+}}, \quad \text{Re}[s]\geq 0.
\end{equation}

Using analytical tools, the average free energy yields
\begin{equation}
F_{q}=\sum _{k=1}^{\infty}\frac{(-1)^{k}a^{k}}{k!k}\mathbb{E}[Z^{k}]+(\ln (a)+\gamma_{e})-R(a),
\label{fq}
\end{equation}
where $\gamma_{e}$ is Euler's constant and
\begin{equation}
|R(a)|\leq \frac{1}{Z(0)a}e^{-Z(0)a},
\end{equation}
with $a$ being an arbitrary dimensionless constant.

\section{Distributional zeta function performance measures}
\label{dzfpm}

In this section, we present the main contribution of this paper. We present the detailed construction of the new set of SPM composed of i) the generalized field network partition funcion; ii) the DZF of a field network; iii) the expected value of the replica field network partition function; and iv) the quenched free energy of a field network. We examine the conditions that they must accomplish to be a SPM and study the physical implications of our interpretations over each object.

\subsection{The generalized field network partition funcion}
We start from the usual Lagrangian density for a complex scalar field $\psi$
\begin{equation}
\mathcal{L}=i\psi ^{*}\frac{\partial \psi}{\partial t}-\frac{1}{2m}\nabla \psi ^{*}\nabla \psi -U\psi ^{*}\psi ,
\end{equation}
where $\nabla$ is the usual gradient operator and $U$ a potential energy function.

The partition function $Z$ is then
\begin{eqnarray}
Z &=& \int d[\psi]\exp \left[ i \int dt \right.
\nonumber \\
&\times & \int d^{d-1}x \left. \left( i\psi ^{*}\frac{\partial \psi ^{*}}{\partial t} - \nabla \psi ^{*}\nabla \psi -V\psi ^{*}\psi \right) \right] .
\end{eqnarray}

Now, we shall move to an Euclidean $d$-dimensional space and a real scalar field $\varphi (x)$. In this case, the partition function takes the form (after an integration by parts)
\begin{equation}
Z=\int d[\varphi] \exp \left[ -\frac{1}{2}\int d^{d}x \varphi(x)\left(- \Delta+U\right) \varphi (x) \right] , 
\label{zvar11}
\end{equation}
where $\Delta$ denotes the Laplacian differential operator in $\mathbb{R}^{d}$. Notice the difference between the Laplacian matrix (\ref{laplacianmatrix000}) and the Laplacian operator $\Delta$ which will be related as follows.
 
Following the same procedure in~\cite{wang5}, we stablish the following identity
\begin{equation*}
-\Delta - U = \tilde{L}=D^{-1/2}LD^{-1/2} .
\end{equation*}

Then, Expresion (\ref{zvar11}) with a disorder source coupled to the fields $h(x)$ yields 
\begin{eqnarray}
Z[\tilde{L},h]= \int d[\varphi] \exp \left[ \frac{1}{2}\int d^{d}x \varphi (x) \tilde{L} \varphi (x) \right.
\nonumber \\
- \left. \int d^{d}x h(x)\varphi (x) \right] ,
\label{z1}
\end{eqnarray}
where $h(x)$ is defined by (\ref{prob1}). Expression (\ref{z1}) is the functional partition function in terms of the normalized Lapacian $\tilde{L}$ of a graph. 

Defining $\varphi _{i}(x)$ as the $i$-th eigenfunction of $\tilde{L}$ with eigenvalue $\tilde{\lambda} _{i}$, we can define a partition function for each eigenfunction as follows (see Appendix \ref{appendix1} to explore the procedure that allows us to do that),
\begin{eqnarray}
Z_{i}[\tilde{L},h]= \int d[\varphi _{i}] \exp \left[ -\frac{1}{2}\int d^{d}x \varphi _{i}(x)\tilde{L}\varphi _{i}(x) \right.
\nonumber \\
- \left. \int d^{d}x h(x)\varphi _{i}(x)\right] .
\end{eqnarray}

Since we are dealing with eigenfunctions of $\tilde{L}$, the above expression yields
\begin{eqnarray}
Z_{i}[\tilde{L},h]= \int d[\varphi] \exp \left[ -\frac{1}{2}\tilde{\lambda} _{i}\int d^{d}x (\varphi _{i}(x))^{2} \right.
\nonumber \\
-\left. \int d^{d}x h(x)\varphi _{i}(x)\right] .
\end{eqnarray}

Now, integrating over the disorder field $Z_{i}[\tilde{L}]=\int d[h]P(h)Z_{i}[\tilde{L},h]$, we have, after perform the Gaussian integrals,
\begin{equation*}
Z_{i}[\tilde{L}]= \sqrt{2\pi \sigma}\int d[\varphi _{i}]\exp \left[ -\frac{1}{2}(\sigma + \tilde{\lambda} _{i})\int d^{d}x(\varphi (x))^{2}\right] .
\end{equation*}

The last expression finally yields

\begin{equation}
Z_{i}[\tilde{L}]\equiv Z_{i}(\sigma , \tilde{\lambda} _{i})= 2\pi \sqrt{\frac{\sigma}{\sigma +\tilde{\lambda} _{i}}} .
\end{equation}

We can move to the domain of the eigenvalues of (\ref{laplacianmatrix000}). Thus, we define the first new SPM as follows
\begin{equation}
\Phi _{Z}(\lambda _{2},...,\lambda _{k};\sigma)=\sum _{i=2}^{k} Z_{i}(\lambda _{i},\sigma ) =2\pi \sum _{i=2}^{k}\sqrt{\frac{\sigma}{\sigma +\lambda _{i}}} .
\label{phiz1}
\end{equation}

Now, we have to show that the spectral function $\Phi _{Z}:\mathbb{R}^{k-1}_{+}\rightarrow \mathbb{R}$ is a SPM. Then we have to show that (\ref{phiz1}) is a decreasing convex function (see Theorem A. 3. in Appendix \ref{appendix2}). We calculate $v^{T}\nabla ^{2} _{\lambda}Z_{i}(\sigma , \lambda _{i}) v$ for any $v\in \mathbb{R}^{k-1}$ and show that it is a real positive value, being $\nabla ^{2}_{\lambda}Z_{i}(\sigma , \lambda _{i})=\text{diag}(\partial ^{2} Z_{2}/ \partial \lambda _{2}^{2},..., \partial ^{2} Z_{k}/ \partial \lambda _{k}^{2})$. From (\ref{phiz1}), we have
\begin{equation*}
v^{T}\nabla ^{2} _{\lambda}Z_{i}(\sigma , \lambda _{i}) v = \frac{3}{4} \sum _{j=1}^{k-1}\frac{\sigma ^{1/2}}{(\sigma + \lambda _{j})^{5/2}}v_{j}^{2} .
\end{equation*}

Therefore (\ref{phiz1}) is a convex and decreasing function; thus it is a SPM.

In virtue of Distributional zeta function of a graph
\begin{equation}
\Phi (s,\tilde{L})=\int d[h]P(h)\frac{1}{Z^{s}[\tilde{L},h]}
\end{equation}
we can evaluate a function

\begin{equation}
\Phi _{Z}(\lambda _{2},...,\lambda _{k};\sigma , p)=2\pi \sum _{i=2}^{k}\left[ \frac{\sigma}{\sigma +\lambda _{i}}\right] ^{\frac{-p+1}{2p}} .
\label{phiz2}
\end{equation}

It is easy to show that (\ref{phiz2}) is a SPM if $-1\leq p \leq 1$.

\subsection{The expected value of the replica field network partition function}

In this case, we use the Laplacian matrix $L(k)$ where the $k$ dependency must be taken as the number of replica fields $\varphi (x)$ acting as nodes. From (\ref{fq}), we can observe that it is necessary to evaluate $Z^{k}$ and its expected value $\mathbb{E}[Z^{k}]$. Beginning with $Z^{k}$, we have the following expression
\begin{equation}
(Z(h))^{k}=\int \prod _{i=1}^{k}d[\varphi _{i}]\exp \left(-\sum _{j=1}^{k}S(\varphi _{i},h)\right) .
\end{equation}

Moreover, using the probability distribution of the disorder (\ref{prob1}), after integrating over the disorder the generic replica partition function yields
\begin{equation}
\mathbb{E}[Z^{k}]=\int \prod _{i=1}^{k}d[\varphi _{i}]\exp (-S_{eff}(\varphi _{i})) ,
\label{exv1}
\end{equation}
being $S_{eff}(\varphi _{i})$ the effective action given by
\begin{eqnarray}
S_{eff}(\varphi _{i})&=&\, \frac{1}{2}\sum _{i,j=1}^{k}\int d^{d}x\int d^{d}y \varphi _{i}(x)D_{ij}(x-y)\varphi _{j}(y) 
\nonumber \\
& &\, +\frac{\lambda _{0}}{4!}\sum _{i=1}^{k}\int d^{d}x\varphi _{i}^{4}(x),
\label{seff1}
\end{eqnarray}
where
\begin{equation}
D_{ij}(x-y)=(\delta _{ij}(-\Delta +m_{0}^{2})-\sigma)\delta ^{d}(x-y),
\label{d11}
\end{equation}
here, $\Delta$ denotes the Laplacian operator in $\mathbb{R}^{d}$, as before. Expressions (\ref{exv1}), (\ref{seff1}) and (\ref{d11}) are analogous to a Euclidean field theory for $k$ interacting replica fields. Expression (\ref{exv1}), with an external source, can be taken as the generating functional of the correlation functions of the model. Using a statistical mechanics concept, it is called a replica partition function.

We can see that disorder fields in this replica scenario define interactions between the replica fields. Let us supposse that the disorder has information about the new interaction between the fields; in this sense we encode the interaction in the disorder parameter $\sigma$ as $\sigma \rightarrow a_{ij}\sigma$ being $a_{ij}$ the adjacency matrix elements.

The expected value of the replica generalized partition function yields, 
\begin{eqnarray}
\mathbb{E}[Z^{k}[L(k)]]=\int \prod _{i=1}^{k}d[\varphi _{i}]\exp \left(-\frac{1}{2}\sum _{\mu ,\nu = 1}^{k} \right. 
\nonumber \\
\left. \int d^{d}x \int d^{d}y \varphi _{\mu}(\delta _{\mu \nu}\l _{\mu \nu}-\sigma a_{\mu \nu})\delta ^{d}(x-y) \varphi _{\nu}\right) .
\end{eqnarray}

Rewriting the above expression we have

\begin{eqnarray}
\mathbb{E}[Z^{k}[L(k)]]=\int \prod _{i=1}^{k}d[\varphi _{i}]\exp \left(-\frac{1}{2}\sum _{\mu ,\nu = 1}^{k} \right.
\nonumber \\
 \left. \int d^{d}x \int d^{d}y \varphi _{\mu}G_{\mu \nu}(k,\sigma)\delta ^{d}(x-y) \varphi _{\nu}\right)  ,
\label{expec11}
\end{eqnarray}

where matrix $G(k,\sigma)$ is defined as
\begin{equation}
G(k,\sigma)= \sigma L(k)+(1-\sigma)D(k) .
\end{equation}

Using the well-known results for Gaussian integrals, we have that the result of (\ref{expec11}) can be written as
\begin{eqnarray}
\mathbb{E}[Z^{k}[L(k)]]&=& \sqrt{\frac{(2\pi)^{k}}{\det[ G(k,\sigma)] }} 
\nonumber \\
 &=& \sqrt{\frac{(2\pi)^{k}}{\sum _{\chi \in S_{k}} \text{sgn}(\chi)\prod _{i=1}^{k}G_{i,\chi _{i}}(k,\sigma)}} .
\end{eqnarray}
where the sum is computed over all permutations $\chi$ of the set $\{1, 2,..., k\}$ and $S_{k}$ is the set of all such permutations (also known as the symmetric group on $k$ elements).

To obtain the explicit dependence with the Laplacian matrix elements, we have
\begin{equation}
\mathbb{E}[Z^{k}[L(k)]]= \sqrt{\frac{(2\pi)^{k}}{\sum _{\chi \in S_{k}} \text{sgn}(\chi)\prod _{i=1}^{k}[\sigma l_{i,\chi _{i}}+(1-\sigma)d_{i,\chi _{i}}]}} .
\label{expec12}
\end{equation}

In order to show that (\ref{expec12}) is a SPM, we can express the last expression in its spectral form as 
\begin{equation}
\Phi _{\mathbb{E}[Z^{k}[\lambda _{i}(k)]]} = \sqrt{\frac{(2\pi)^{k}}{\prod _{i=1}^{k}[(1-\sigma)d_{ii}+\sigma \lambda _{i}(k)]}} .
\label{expect13}
\end{equation}

We need to show that $\nabla _{\lambda} ^{2}\Phi _{\mathbb{E}[Z^{k}[\lambda _{i}(k)]]} \preceq 0$, where the Hessian of (\ref{expect13}) is given by
\begin{equation}
\frac{\partial ^{2}\Phi _{\mathbb{E}[Z^{k}[\lambda _{i}(k)]]}}{\partial \lambda _{j}\partial \lambda _{i}}=\frac{\sigma ^{2}}{2}\Phi _{\mathbb{E}[Z^{k}[\lambda _{i}(k)]]}\left[\frac{1}{2\beta _{i}\beta _{j}}+\frac{\delta _{ij}}{\beta _{i}^{2}} \right],
\end{equation}
being
\begin{equation}
\beta _{i}=(1-\sigma)d_{ii}+\sigma \lambda _{i}.
\label{beta11}
\end{equation}

As before, to verify $\nabla _{\lambda}^{2}\Phi _{\mathbb{E}[Z^{k}[\lambda _{i}(k)]]} \preceq 0$ we must show that $v^{T}\nabla _{\lambda} ^{2}\Phi _{\mathbb{E}[Z^{k}[\lambda _{i}(k)]]} v \leq 0$ for all vectors $v\in \mathbb{R}^{k}$. We have that

\begin{eqnarray}
v^{T}\nabla _{\lambda}^{2}\Phi _{\mathbb{E}[Z^{k}[\lambda _{i}(k)]]} v = & &\frac{3\sigma ^{2}}{4}\Phi _{\mathbb{E}[Z^{k}[\lambda _{i}(k)]]}v^{t}\Lambda v 
\nonumber \\
&+&\frac{\sigma ^{2}}{4}\Phi _{\mathbb{E}[Z^{k}[\lambda _{i}(k)]]}v^{T}\Gamma v ,
\label{expect14}
\end{eqnarray}
being $\Lambda$ a diagonal positive-definite matrix defined by
\begin{equation}
\Lambda = \begin{pmatrix}
\beta _{1}^{-2} & 0 & \cdots & 0 \\
0 & \beta _{2}^{-2} & \cdots & 0 \\
\vdots & \vdots & \ddots & \vdots \\
0 & 0 & \cdots & \beta _{k}^{-2}
\end{pmatrix} ,
\end{equation}
and $\Gamma$ is a hollow symmetric nonnegative (HSN) matrix defined by
\begin{equation}
\Gamma = \begin{pmatrix}
0 & (\beta _{1}\beta _{2})^{-1} & \cdots & (\beta _{1}\beta _{k})^{-1} \\
(\beta _{2}\beta _{1})^{-1} & 0 & \cdots & (\beta _{2}\beta _{k})^{-1} \\
\vdots & \vdots & \ddots & \vdots \\
(\beta _{k}\beta _{1})^{-1} & (\beta _{k}\beta _{2})^{-1} & \cdots & 0
\end{pmatrix} .
\label{gamma1111}
\end{equation}

By using Ramsey-based theorems~\cite{ramsey1, lamaison1, causey1, choi1, charles1, farber1, johnson1} we can extract relations for the eigenvalues of (\ref{gamma1111}) (see Appendix \ref{appendix3} for further development). With these relations we will have two scenarios. Firstly, when $v^{T}\Gamma v \geq 0$ or $v^{T}\Gamma v \leq 0$ and $|v^{T}\Gamma v| \leq 3v^{T}\Lambda v$, the function (\ref{expec12}) is convex. Thus (\ref{expect13}) is a SPM.

On the other hand, if $v^{T}\Gamma v \leq 0$ and $3v^{T}\Lambda v \leq |v^{T}\Gamma v|$ the function (\ref{expec12}) is concave. Now with the following decreasing convex function
\begin{equation*}
h_{a}(x) = x^{\frac{-p+1}{p}}\quad \text{for all}\quad 2\leq p \leq \infty
\end{equation*}

We have that 
\begin{eqnarray}
\Phi _{\mathbb{E}[Z^{k}]_{p}}(\lambda _{1},...,\lambda _{k};\sigma ,p)&=&h_{a}(\mathbb{E}[Z^{k}[\lambda _{i}(k)]]) 
\nonumber \\
&=& (\mathbb{E}[Z^{k}[\lambda _{i}(k)]]) ^{\frac{-p+1}{p}}
\label{expect22}
\end{eqnarray}
is convex, then (\ref{expect22}) is a SPM.

In order to apply the growing Algorithms 1 and 2 we define the matrix elements $(\nabla _{L}\mathbb{E}[Z^{k}[L(k)]])_{\mu \nu}$ as
\begin{equation*}
(\nabla _{L}\mathbb{E}[Z^{k}[L(k)]])_{\mu \nu} = \frac{\partial \mathbb{E}[Z^{k}[L(k)]]}{\partial l _{\mu \nu}}
\end{equation*}
\begin{equation}
= \frac{\partial }{\partial l _{\mu \nu}}\left( \sqrt{\frac{(2\pi)^{k}}{\sum _{\chi \in S_{k}} \text{sgn}(\chi)\prod _{i=1}^{k}[\sigma l_{i,\chi _{i}}+(1-\sigma)d_{i,\chi _{i}}]}} \right) .
\end{equation}

After some algebraic steps, and applying some matrix determinant properties, we have
\begin{equation}
(\nabla _{L}\mathbb{E}[Z^{k}[L(k)]])_{\mu \nu} = -\frac{(\mathbb{E}[Z^{k}[L(k)]])^{3}}{2(2\pi)^{k}}\mathcal{L} _{\mu \nu}(k)
\end{equation}
being the matrix elements $\mathcal{L} _{\mu \nu}(k)$ defined by
\begin{widetext}
\begin{equation}
\mathcal{L} _{\mu \nu}(k) = \sum _{i_{1},...,i_{k}=1}^{k}\left\lbrace \epsilon _{i_{1},...,i_{k}}\sum _{\alpha = 1}^{k}\left[ \sigma \delta _{\alpha , \mu} \delta _{i_{\alpha},\nu}\prod _{\beta \neq \alpha} \left[  (1-\sigma) d_{\beta ,i_{\beta}}+\sigma l _{\beta ,i_{\beta}} \right] \right] \right\rbrace
\end{equation}
\end{widetext}
where $\delta _{\alpha ,\beta}$ is the usual Kronecker delta and $\epsilon _{i_{1},...,i_{k}}$ is the totally antisymmetric Levi-Civita symbol.

Finally, let us evaluate the logarithm of the expected value. We have that
\begin{equation*}
\log (\mathbb{E}[Z^{k}[L(k)]])=\frac{k}{2}\log (2\pi)-\log \left( \sum _{\gamma \in S_{k}}\text{sgn}(\gamma)\prod _{i=1}^{k}a_{i,\gamma _{i}}\right) .
\end{equation*}

Its spectral representation is given by
\begin{equation}
\Phi _{\log(\mathbb{E}[Z^{k}])}(\lambda _{1},...,\lambda _{k};\sigma )=\frac{k}{2}\log (2\pi)-\sum _{i=1}^{k}\log[\beta _{i}] ,
\label{log11}
\end{equation} 
remembering that $\beta _{i}\equiv \beta_{i}(\lambda _{i})$ is defined by (\ref{beta11}). To show that function (\ref{log11}) is a SPM we show that $v^{T}\nabla ^{2} _{\lambda}\Phi _{\log(\mathbb{E}[Z^{k}])} v \geq 0$ for any $v\in \mathbb{R}^{n}$, this quantity yields
\begin{equation*}
v^{T}\nabla ^{2} _{\lambda}\Phi _{\log(\mathbb{E}[Z^{k}])} v = \sigma ^{2}\sum _{i=1}^{k}\frac{v_{i}^{2}}{[(1-\sigma)d_{ii}+\sigma \lambda _{i}]^{2}} \geq 0 .
\end{equation*}

Therefore (\ref{log11}) is a SPM.
\subsection{The quenched free energy of a field network}

Following the discusion that offers the DZF approach~\cite{svaiter2}, we have that the average free energy is writen as a series of the integer moments of the partition function of the model. It is shown that there exists a spontaneous symmetry breaking mechanism in the disordered model. In order to show this mechanism, we have that the ground state configurations of a field $\varphi (x)$ are defined by the following saddle-point equation
\begin{equation*}
(-\Delta +m_{0}^{2})\varphi _{h}(x)+\frac{\lambda _{0}}{3!}\varphi _{h}^{3}(x)=h(x)
\end{equation*}
where $\varphi _{h}(x)$ denotes the field defined for a particular configuration of disorder. After integrating out the disorder field in a generic replica partition function, the saddle-point equation yields
\begin{equation*}
(-\Delta +m_{0}^{2})\varphi _{i}(x)+\frac{\lambda _{0}}{3!}\varphi _{i}^{3}(x)=\sigma \sum _{j=1}^{k}\varphi _{j}(x) .
\end{equation*}

Applying the replica symmetric ansatz, we have
\begin{equation}
(-\Delta +m_{0}^{2}-k\sigma)\varphi _{i}(x)+\frac{\lambda _{0}}{3!}\varphi _{i}^{3}(x)= 0 .
\label{saddle111}
\end{equation}

In this approach, we must take into account all replica partition functions contributing to the average free energy. Assuming $m_{0}^{2}>0$, a critical $k_{c}$ is defined as $k_{c}=\lfloor m_{0}^{2} /\sigma \rfloor$ where $\lfloor x \rfloor$ denotes the integer part of $x$. For $m_{0}^{2}\geq \sigma$, $m_{0}^{2}-k\sigma \geq 0$ is satisfied as $k_{c}\leq k$. From (\ref{saddle111}), in this situation, each replica field fluctuates around the zero value, whichs is understood as the stable equilibrium state. This scenario is different for the contributions where $k_{c}\geq k$. In this situation the replica fields, with $k \leq k_{c}$, still fluctuating around the zero value. Nevertheless it is not an equilibrium state anymore. In the framework of field operators, this means that if we compute the vacuum expectation value of such fields, it does not vanish. This is precisely the schema in which spontaneous symmetry breaking emerges.

For a very large $a$ the dominant contribution of (\ref{fq}) can be written as 
\begin{equation}
F_{q}=\sum _{k=1}^{N}\frac{(-1)^{k}a^{k}}{k!k}\mathbb{E}[Z^{k}[L(k)]] .
\label{free1}
\end{equation}  

Observe that this series representation has two kinds of replica partition functions. For $k\leq k_{c}$, and following Huckel orbital method~\cite{streit1, coulson1} (neglecting self-interactions) we have
\begin{equation}
\mathbb{E}^{(1)}[Z^{k}[L(k)]]=\int \prod _{i=1}^{k}d[\varphi _{i}]\exp (-S_{eff}(\varphi _{i}))
\end{equation}
where
\begin{equation}
S_{eff}(\varphi _{i})=\frac{1}{2}\sum _{i,j=1}^{k}\int d^{d}x \int d^{d}y \varphi _{i}(x)B_{ij}(m_{0},\sigma ;x-y)\varphi _{j}(x)
\end{equation}
being the operator matrix elements $B_{ij}$ defined by
\begin{equation}
B_{ij}(m_{0},\sigma ;x-y)=(\delta _{ij}(l_{ij}+m_{0}^{2})-a_{ij}\sigma )\delta^{d}(x-y) .
\end{equation}

For $k_{c}<k \leq N$, the replica partition function $\mathbb{E}^{(2)}[Z^{k}[L(k)]]$ is
\begin{equation}
\mathbb{E}^{(2)}[Z^{k}[L(k)]]=\int \prod _{j=1}^{k}d[\phi _{j}]\exp (-S_{eff}(\phi _{j}))
\end{equation}
where
\begin{eqnarray}
S_{eff}(\phi _{i})&=&\frac{1}{2}\sum _{i,j=1}^{k}\int d^{d}x 
\nonumber \\
&\times & \int d^{d}y \phi _{i}(x)C_{ij}(m_{0},\sigma , N;x-y)\phi _{j}(x)
\end{eqnarray}
being the operator matrix elements $C_{ij}$ defined by
\begin{equation}
C_{ij}(m_{0},\sigma , N ;x-y)=(\delta _{ij}(l_{ij}+3\sigma N-2m_{0}^{2})-a_{ij}\sigma )\delta^{d}(x-y) .
\end{equation}

Thus, we can reewrite (\ref{free1}) as
\begin{eqnarray}
F_{q}&=& \sum _{k=1}^{k_{c}} \frac{(-1)^{k}a^{k}}{k!k}\mathbb{E} ^{(1)}[Z^{k}[L(k)]] 
\nonumber \\
&+& \sum _{k=k_{c}+1}^{N} \frac{(-1)^{k}a^{k}}{k!k}\mathbb{E} ^{(2)}[Z^{k}[L(k)]] .
\label{free2}
\end{eqnarray}

Each term of the series expansion (\ref{free2}) can be interpreted as a subgraph contribution of $k$ interacting fields of the total graph of $N$ interacting fields. Therefore, we shall now examine when (\ref{free1}) could be a SPM.

Following the analysis of the previous subsection we shall define the matrix set of interest as follow,

\begin{equation}
\Lambda ^{(1)} = \begin{pmatrix}
\varsigma _{1}^{-2} & 0 & \cdots & 0 \\
0 & \varsigma _{2}^{-2} & \cdots & 0 \\
\vdots & \vdots & \ddots & \vdots \\
0 & 0 & \cdots & \varsigma _{k}^{-2}
\end{pmatrix} ,
\end{equation}
and
\begin{equation}
\Gamma ^{(1)}= \begin{pmatrix}
0 & (\varsigma _{1}\varsigma _{2})^{-1} & \cdots & (\varsigma _{1}\varsigma _{k})^{-1} \\
(\varsigma _{2}\varsigma _{1})^{-1} & 0 & \cdots & (\varsigma _{2}\varsigma _{k})^{-1} \\
\vdots & \vdots & \ddots & \vdots \\
(\varsigma _{k}\varsigma _{1})^{-1} & (\varsigma _{k}\varsigma _{2})^{-1} & \cdots & 0
\end{pmatrix} .
\end{equation}

being
\begin{equation}
\varsigma _{i}=(1-\sigma)d_{ii}+\sigma \lambda _{i}+m_{0}^{2}\delta _{ii} .
\end{equation}

On the other hand

\begin{equation}
\Lambda ^{(2)} = \begin{pmatrix}
\zeta _{1}^{-2} & 0 & \cdots & 0 \\
0 & \zeta _{2}^{-2} & \cdots & 0 \\
\vdots & \vdots & \ddots & \vdots \\
0 & 0 & \cdots & \zeta _{k}^{-2}
\end{pmatrix} ,
\end{equation}
and
\begin{equation}
\Gamma ^{(2)}= \begin{pmatrix}
0 & (\zeta _{1}\zeta _{2})^{-1} & \cdots & (\zeta _{1}\zeta _{k})^{-1} \\
(\zeta _{2}\zeta _{1})^{-1} & 0 & \cdots & (\zeta _{2}\zeta _{k})^{-1} \\
\vdots & \vdots & \ddots & \vdots \\
(\zeta _{k}\zeta _{1})^{-1} & (\zeta _{k}\zeta _{2})^{-1} & \cdots & 0
\end{pmatrix} .
\end{equation}

being
\begin{equation}
\zeta _{i}=(1-\sigma)d_{ii}+\sigma \lambda _{i}+(3\sigma N-2m_{0}^{2})\delta _{ii} .
\end{equation}

Let the eigenvalues of $\Gamma ^{(j)}$ be denoted by $\{ \gamma ^{(j)}_{i} \}$ ($j=1,2$) and assume that they have been arranged in nonincreasing order $\gamma ^{(j)}_{1}\geq \cdots \geq \gamma ^{(j)}_{k}$. Then if $v^{T}\Gamma ^{(j)}v \leq 0$ the eigenvalues must accomplish the following relation
\begin{equation}
0 \leq \gamma ^{(j)}_{1}v_{1}^{2} \leq \sum _{i=2}^{k}\gamma ^{(j)}_{i}v_{i}^{2} .
\end{equation}

Otherwise, if $v^{T}\Gamma ^{(j)}v \geq 0$, we have that
\begin{equation}
0 \leq \sum _{i=2}^{k}\gamma ^{(j)}_{i}v_{i}^{2} \leq \gamma ^{(j)}_{1}v_{1}^{2} .
\end{equation}

Then, when $v^{T}\Gamma ^{(j)}v \geq 0$ or $v^{T}\Gamma ^{(j)}v \leq 0$ and $|v^{T}\Gamma ^{(j)} v| \leq 3v^{T}\Lambda ^{(j)}v$, for each $k$ the function (\ref{free2}) is convex. Thus the following function

\begin{eqnarray}
\Phi _{F_{q}}(\lambda _{1},...,\lambda _{k})&=& \sum _{k=1}^{k_{c}} \frac{(-1)^{k}}{k!k}\sqrt{\frac{(2\pi)^{k}}{\prod _{i=1}^{k}[\varsigma _{i}(k)]}}
\nonumber \\
&+&\sum _{k=k_{c}+1}^{N} \frac{(-1)^{k}}{k!k}\sqrt{\frac{(2\pi)^{k}}{\prod _{i=1}^{k}[\zeta _{i}(k)]}}
\end{eqnarray}

is a SPM. On the other hand, if $v^{T}\Gamma ^{(j)}v \leq 0$ and $3v^{T}\Lambda ^{(j)}v \leq |v^{T}\Gamma ^{(j)}v|$ the function (\ref{free2}) is concave. We use the auxiliar convex function $h_{a}(x)$ giving us the following SMP function
\begin{eqnarray}
\Phi _{F_{q}^{(p)}}(\lambda _{1},...,\lambda _{k};p)&=& \sum _{k=1}^{k_{c}} \frac{(-1)^{k}}{k!k}\left[ \frac{(2\pi)^{k}}{\prod _{i=1}^{k}[\varsigma _{i}(k)]} \right] ^{\frac{-p+1}{2p}}
\nonumber \\
&+&\sum _{k=k_{c}+1}^{N} \frac{(-1)^{k}}{k!k}\left[ \frac{(2\pi)^{k}}{\prod _{i=1}^{k}[\zeta _{i}(k)]}\right] ^{\frac{-p+1}{2p}}
\label{energy333}
\end{eqnarray}

Therefore the matrix $\nabla _{L} F_{q}$ is given by
\begin{eqnarray}
\nabla _{L} F_{q} &=& \sum _{k=1}^{k_{c}}\frac{(-1)^{k}}{k!k}\mathbbm{1}^{(k)}\otimes \nabla _{L} \mathbb{E}^{(1)}[Z^{k}[L](k)]
\nonumber \\
&+& \sum _{k=k_{c}+1}^{N}\frac{(-1)^{k}}{k!k}\mathbbm{1}^{(k)}\otimes \nabla _{L} \mathbb{E}^{(2)}[Z^{k}[L](k)]
\end{eqnarray}

where $\mathbbm{1}^{(k)}$ is a $D=N/k$-fractal-dimensional matrix, defined by (\ref{fractal111}), where its unique non-null element is the entry $\mathbbm{1}^{(k)}_{11}=1$, and $\otimes$ is the usual Kronecker product. 

\section{Numerical results}
\label{numericalresults}

\begin{figure}[h]
	\includegraphics[width=0.32\textwidth]{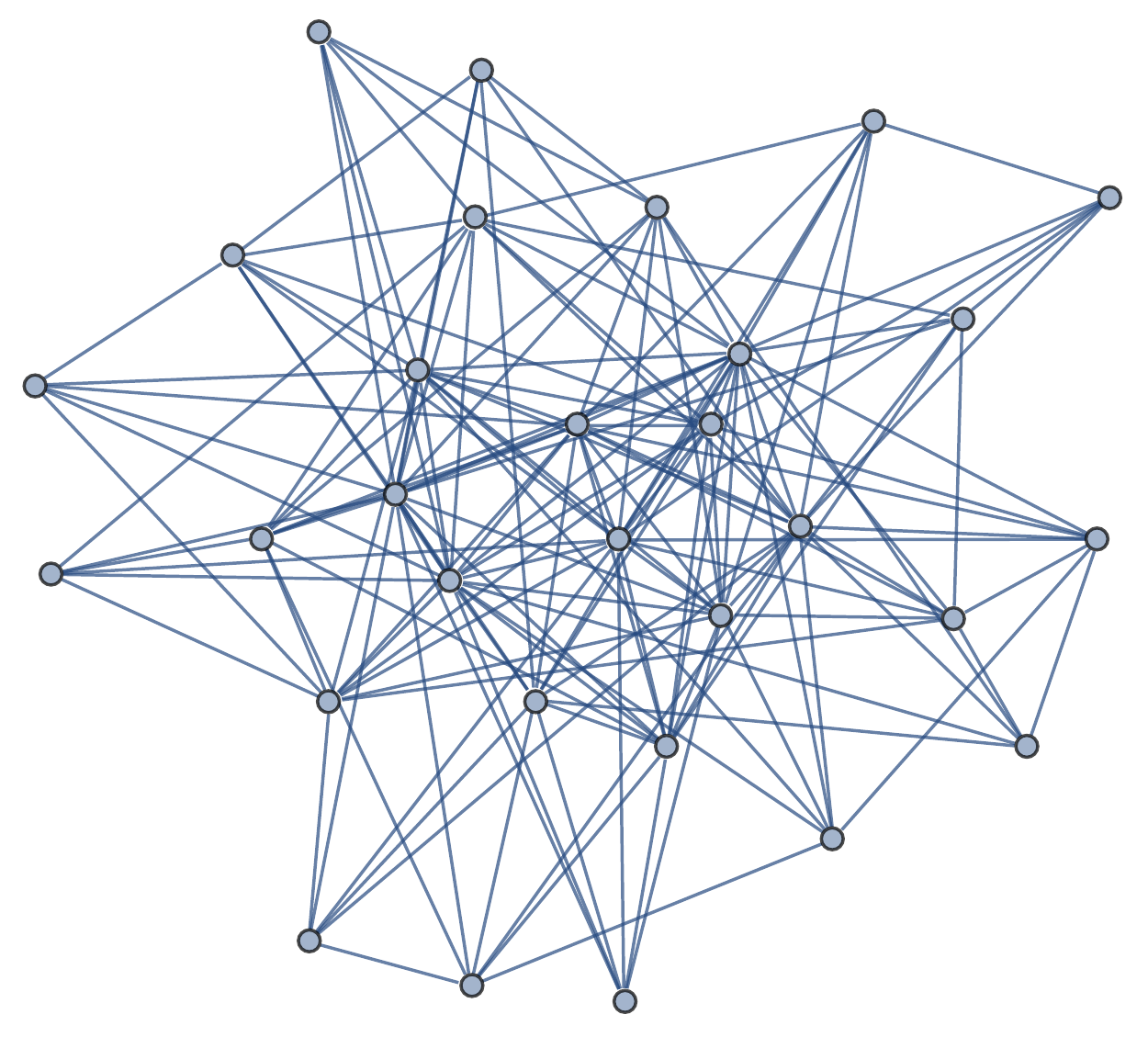}\\
	\caption{Barab\'asi-Albert Network with 30 nodes and 6 edges added at each step.}
	\label{net1}
 \end{figure}

In this section, we explore numerical results to show the advantages of the proposed SPM. We shall work with a Barab\'asi-Albert Network with $k=30$ nodes and 6 edges added at each step; see FIG. (\ref{net1}).

First, we shall study the following quantity
\begin{equation}
\pi _{n}:=\frac{\varrho _{0}-\varrho _{n}}{\varrho _{0}}\times 100
\label{percentage1111}
\end{equation}
that represents the percentage of performance enhancement for all values of design parameter $1\leq n\leq k-1$ (see Appendix \ref{appendix2} for the nature of this quantity, $n$ is the number of new links added).

\begin{figure}[htb]
	\includegraphics[width=.72\columnwidth]{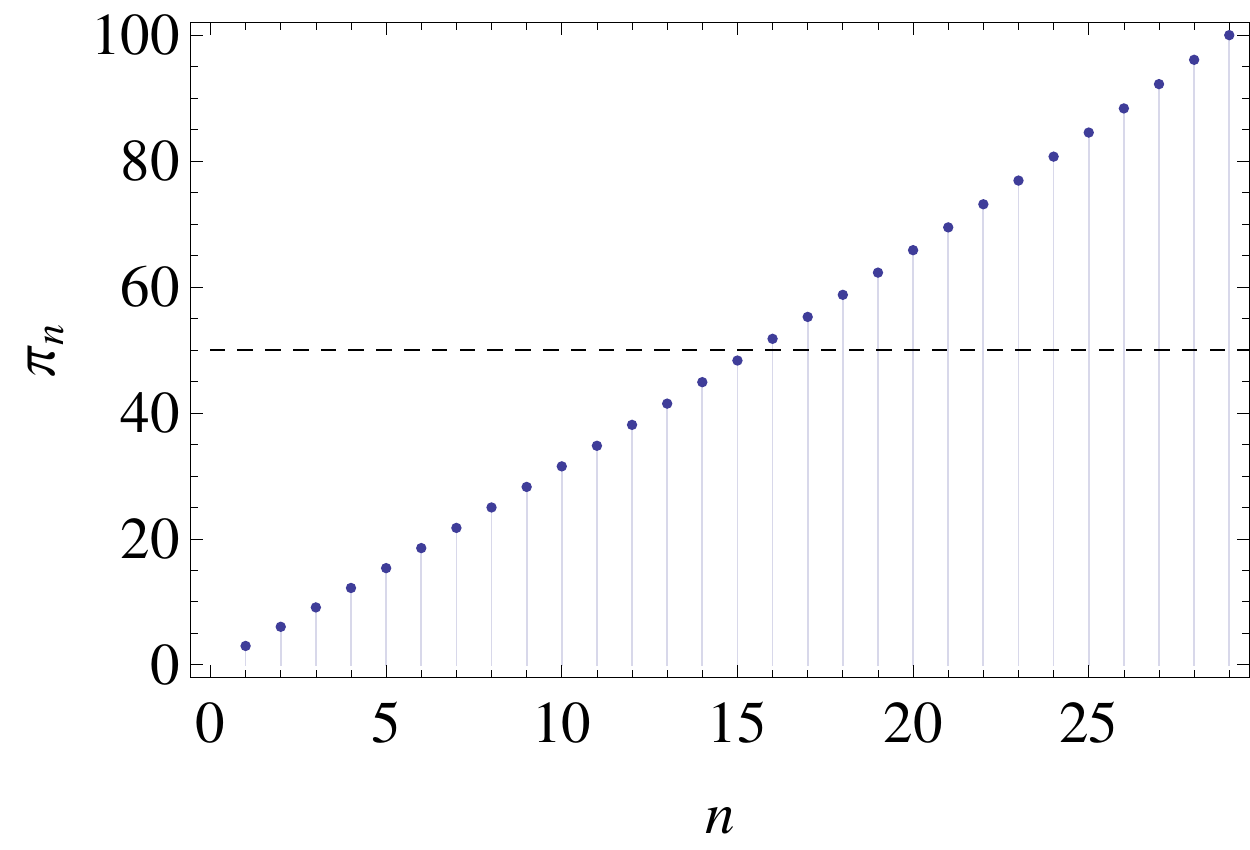}
	\caption{Value of $\pi _{n}$ with respect to the SPM from partition function~(\ref{phiz1}) with $\sigma = 0.1$.}
	\label{figpartition1}
\end{figure}

\begin{figure}[h]
	\centering
	\includegraphics[width=.85\columnwidth]{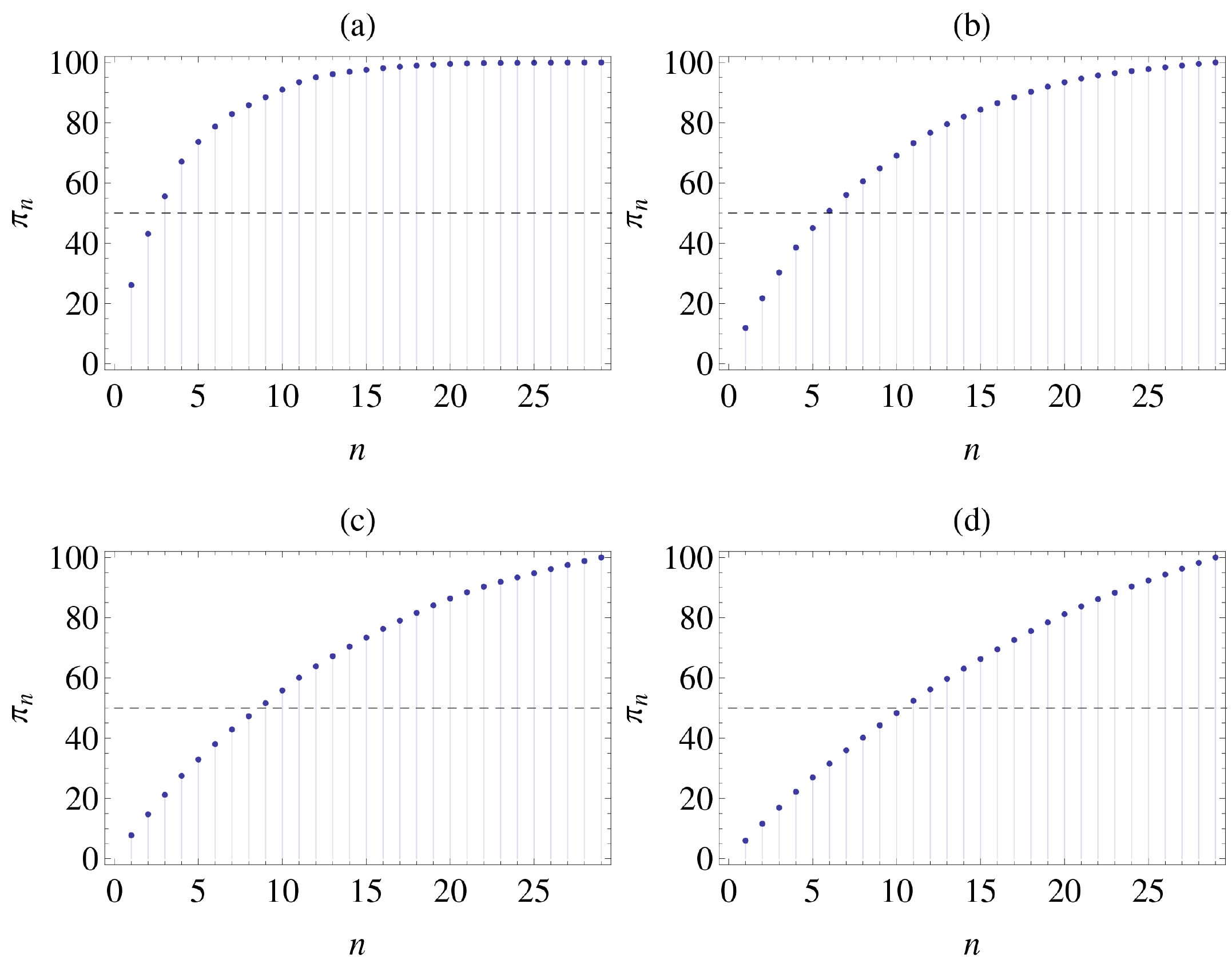}
	\caption{Value of $\pi _{n}$ with respect to $\Phi _{Z}(\lambda _{2},...,\lambda _{n},\sigma = 0.1;p)$ for (a) $p=0.1$, (b) $p=0.2$, (c) $p=0.3$, (d) $p=0.4$.}
	\label{figpartition2}
\end{figure}

\begin{figure}[h]
	\centering
	\includegraphics[width=.79\columnwidth]{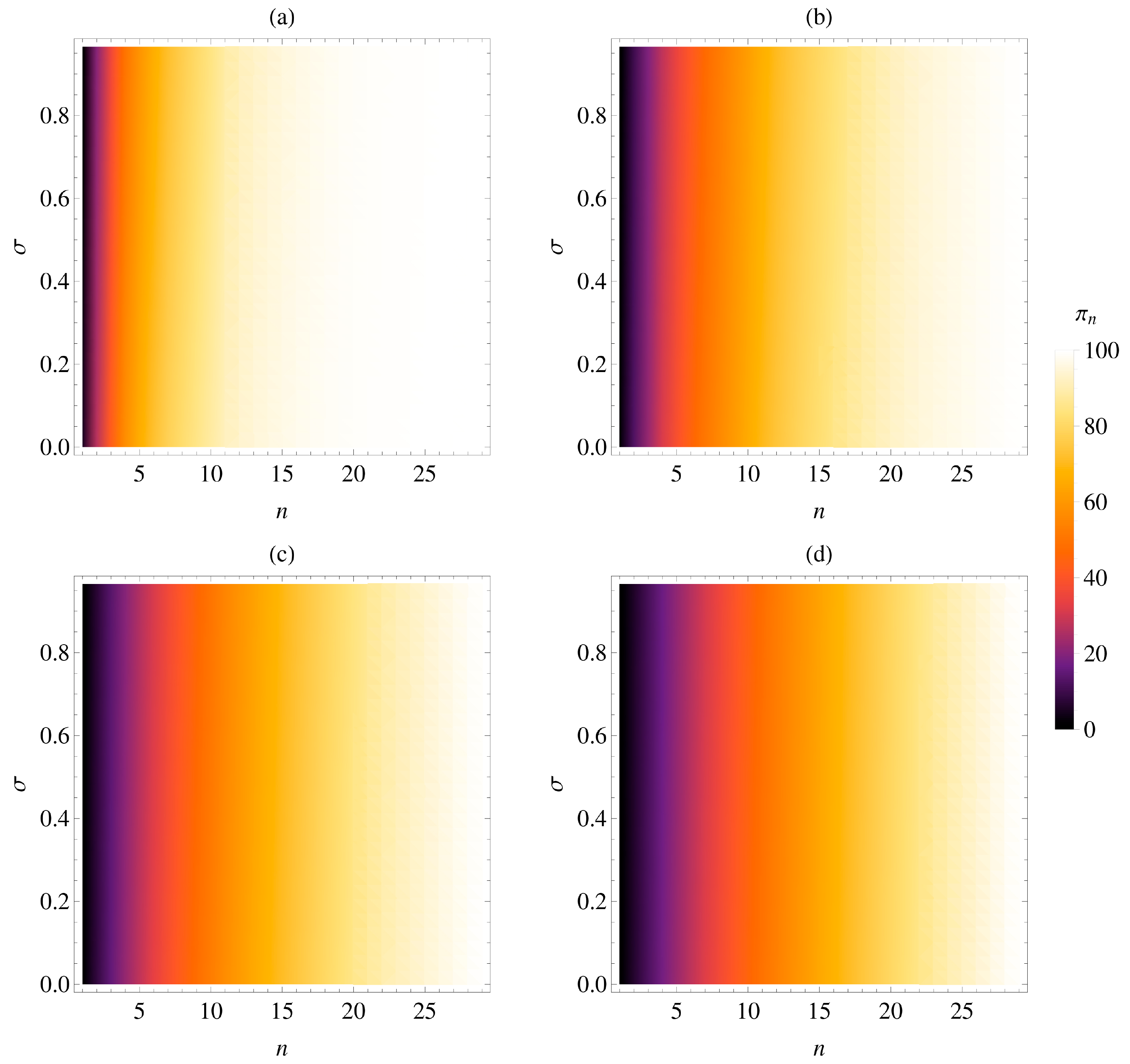}
	\caption{Value of $\pi _{n}$ with respect to $\Phi _{Z}(\lambda _{2},...,\lambda _{n},\sigma;p)$ and the disorder parameter $\sigma$ for (a) $p=0.1$, (b) $p=0.2$, (c) $p=0.3$, (d) $p=0.4$.}
	\label{figpartition3}
\end{figure}

The results for the SPM from partition function (\ref{phiz1}) show that the 50\% performance improvement is achieved by adding 15 and 16 links. In the FIG. \ref{figpartition1} we can observe a similar behavior with respect the SPM reported in~\cite{siami5}. However, something interesting occurs with values of $p<1$. Within these values, we can achieve the 50\% performance improvement by adding less than 9 links. It shows a great advantage in the network synthesis problem. We can improve the performance with a few operations. In FIG. \ref{figpartition2} we can see that for $p=0.1$, $p=0.2$, $p=0.3$, $p=0.4$, we have to add at least 3-4, 5-6, 8-9, 10-11 links, respectively, to achive the 50\% performance improvement. Showing a great advantage respect other known SPM. Now, we may compute the behavior of (\ref{percentage1111}) with respect to the disorder parameter $\sigma$. The results depicted in FIG. \ref{figpartition3} are showing that the low disorder regime will bring us a best performance improvement by adding few links.

\begin{figure}[h]
	\centering
	\includegraphics[width=.85\columnwidth]{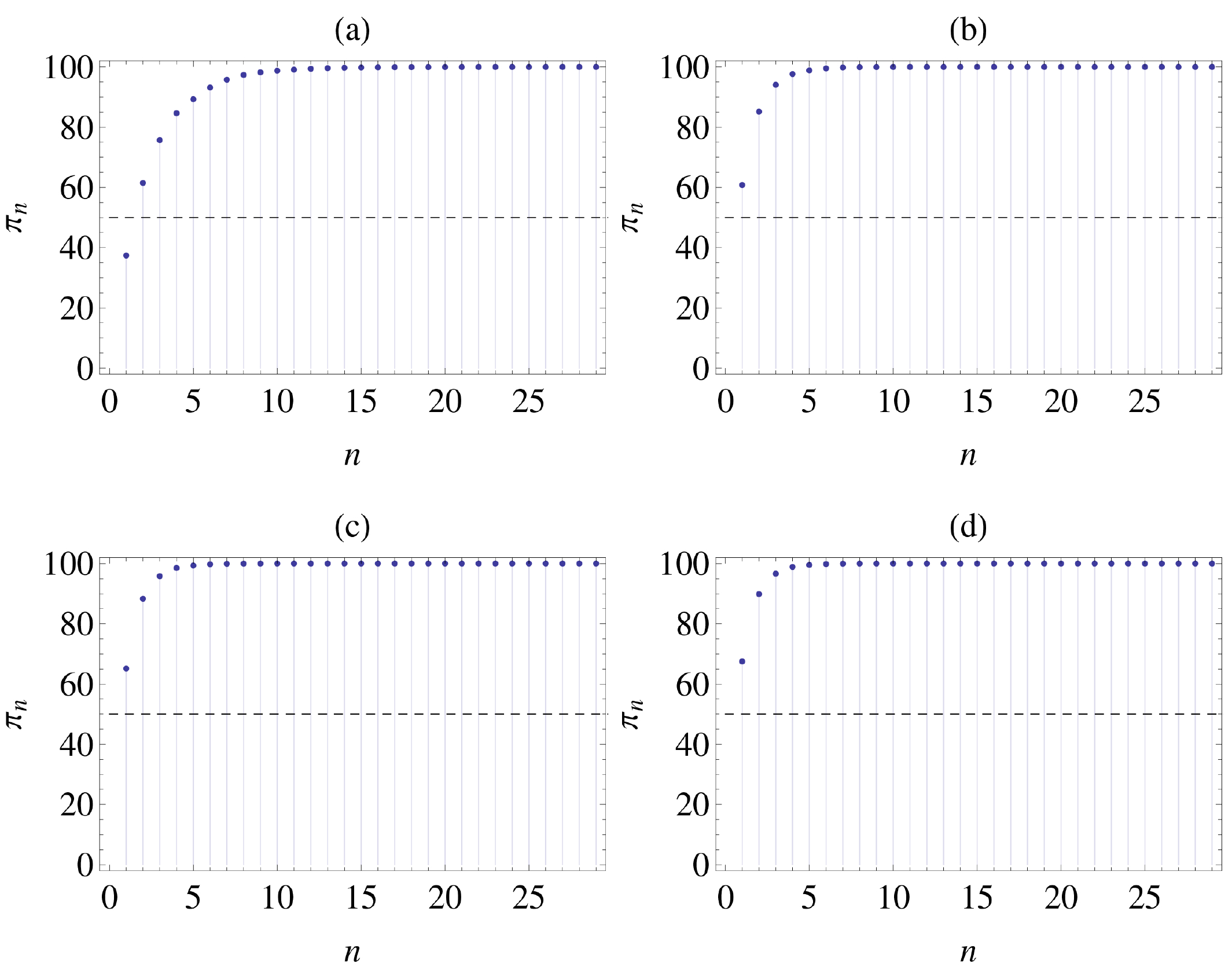}
	\caption{Value of $\pi _{n}$ with respect to $\Phi _{\mathbb{E}[Z^{k}]_{p}}(\lambda _{2},...,\lambda _{n},\sigma =0.1;p)$ for (a) $p=1.5$, (b) $p=3.0$, (c) $p=4.0$, (d) $p=5.0$. }
	\label{figexpected1}
\end{figure}

For the SPM from expected value of the replica generalized partition function (\ref{expect22}), we have that, in contrast with the partition function, for large values of $p$ we observe that the 50\% performance improvement is achieved with less links than the above case. See FIG. \ref{figexpected1} for $p=1.5$, $p=3.0$, $p=4.0$, $p=5.0$.

\begin{figure}[h]
	\centering
	\includegraphics[width=.72\columnwidth]{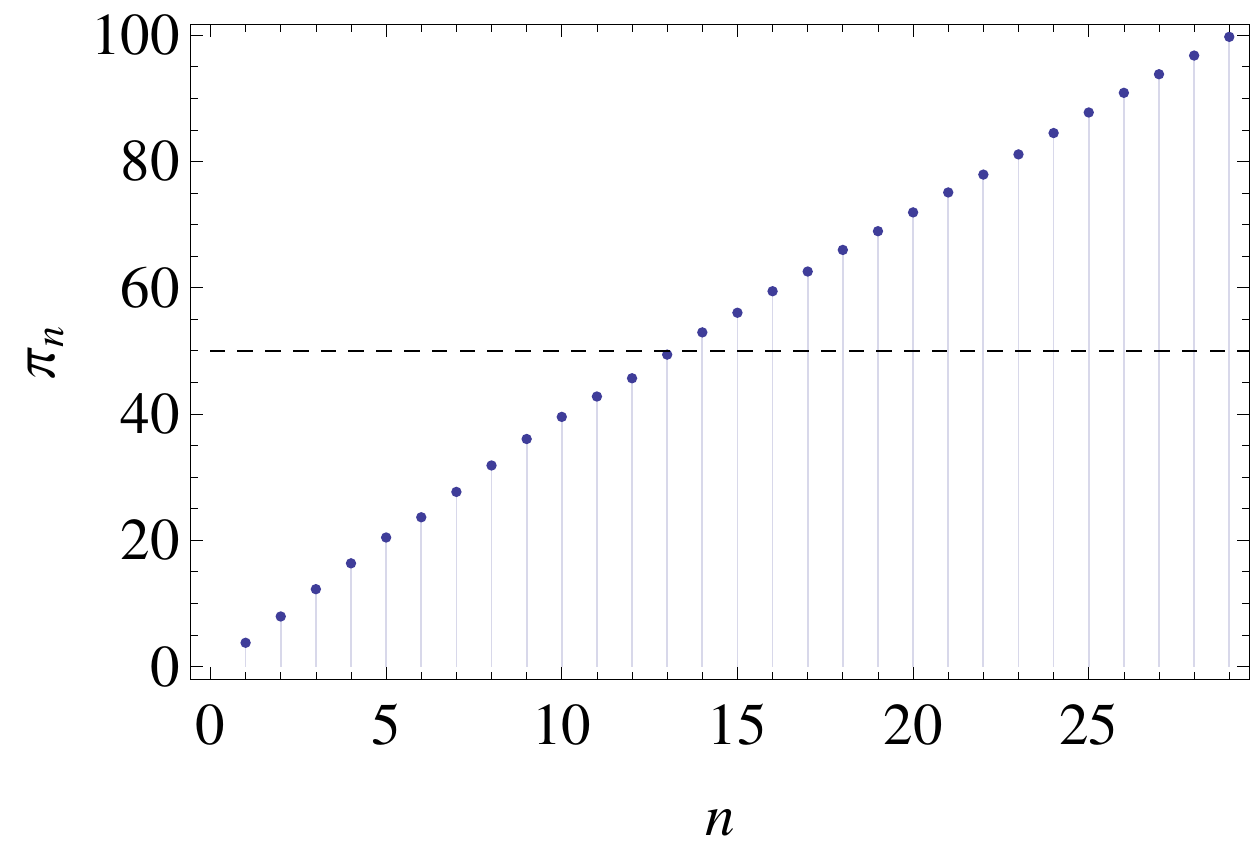}
	\caption{Value of $\pi _{n}$ with respect to $\Phi _{\log(\mathbb{E}[Z^{k}])}$ with $\sigma = 0.1$. }
	\label{figexpectedlog1}
\end{figure}

\begin{figure}[htb]
	\centering
	\includegraphics[width=.72\columnwidth]{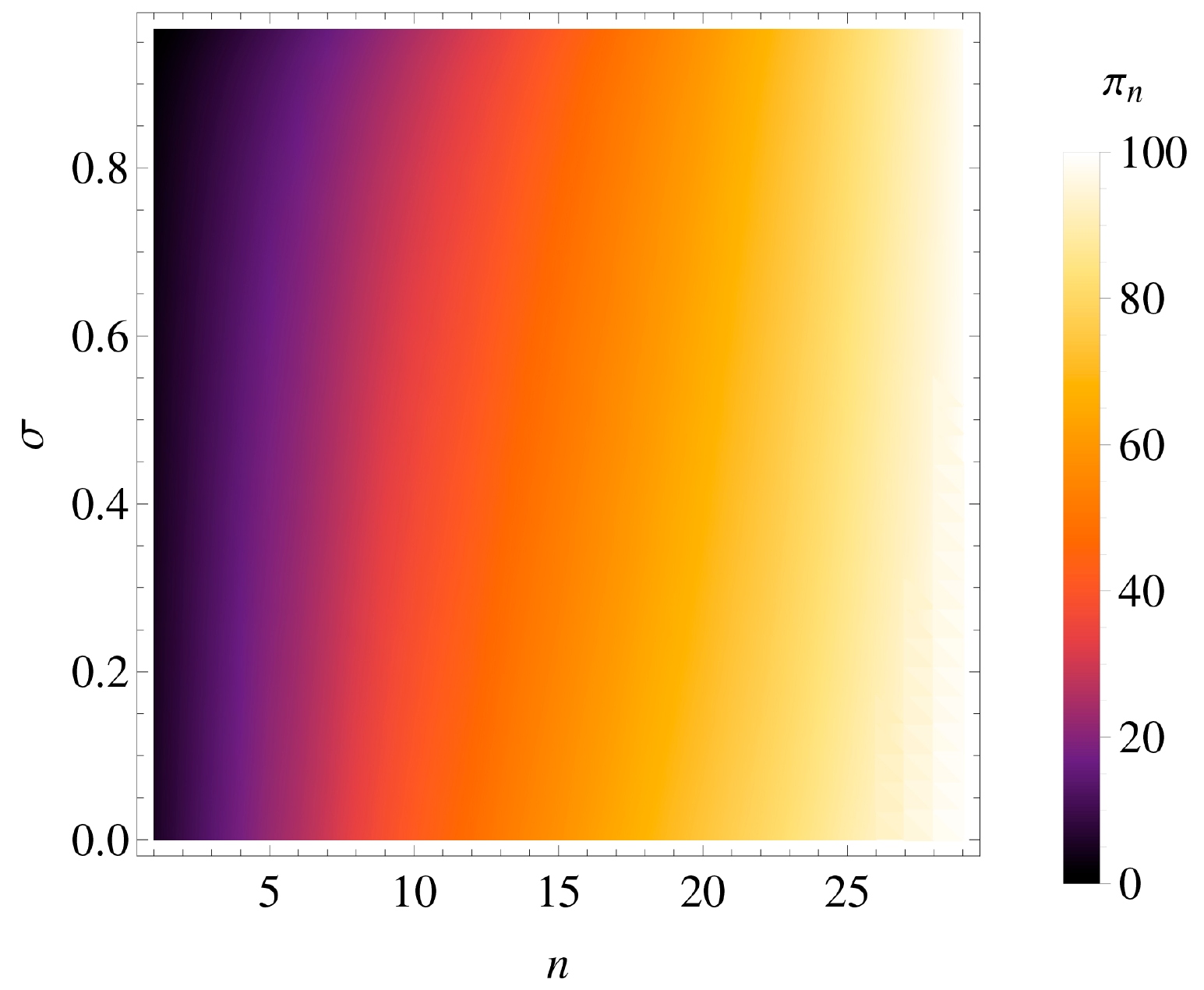}
	\caption{Value of $\pi _{n}$ with respect to $\Phi _{\log(\mathbb{E}[Z^{k}])}$ and $\sigma$.}
	\label{figexpectedlog2}
\end{figure}

However, For the logarithm of expected value $\Phi _{\log(\mathbb{E}[Z^{k}])}$ we observe similar behavior to the known SMP. See FIG. \ref{figexpectedlog1} and FIG. \ref{figexpectedlog2}. In the FIG. \ref{figexpectedlog2}, we can evidence more explicitly the effect of disorder parameter, i.e., in a regime of high disorder is more difficult to reach a optimal performance enhancement with few steps.

\begin{figure}[htb]
	\centering
	\includegraphics[width=.79\columnwidth]{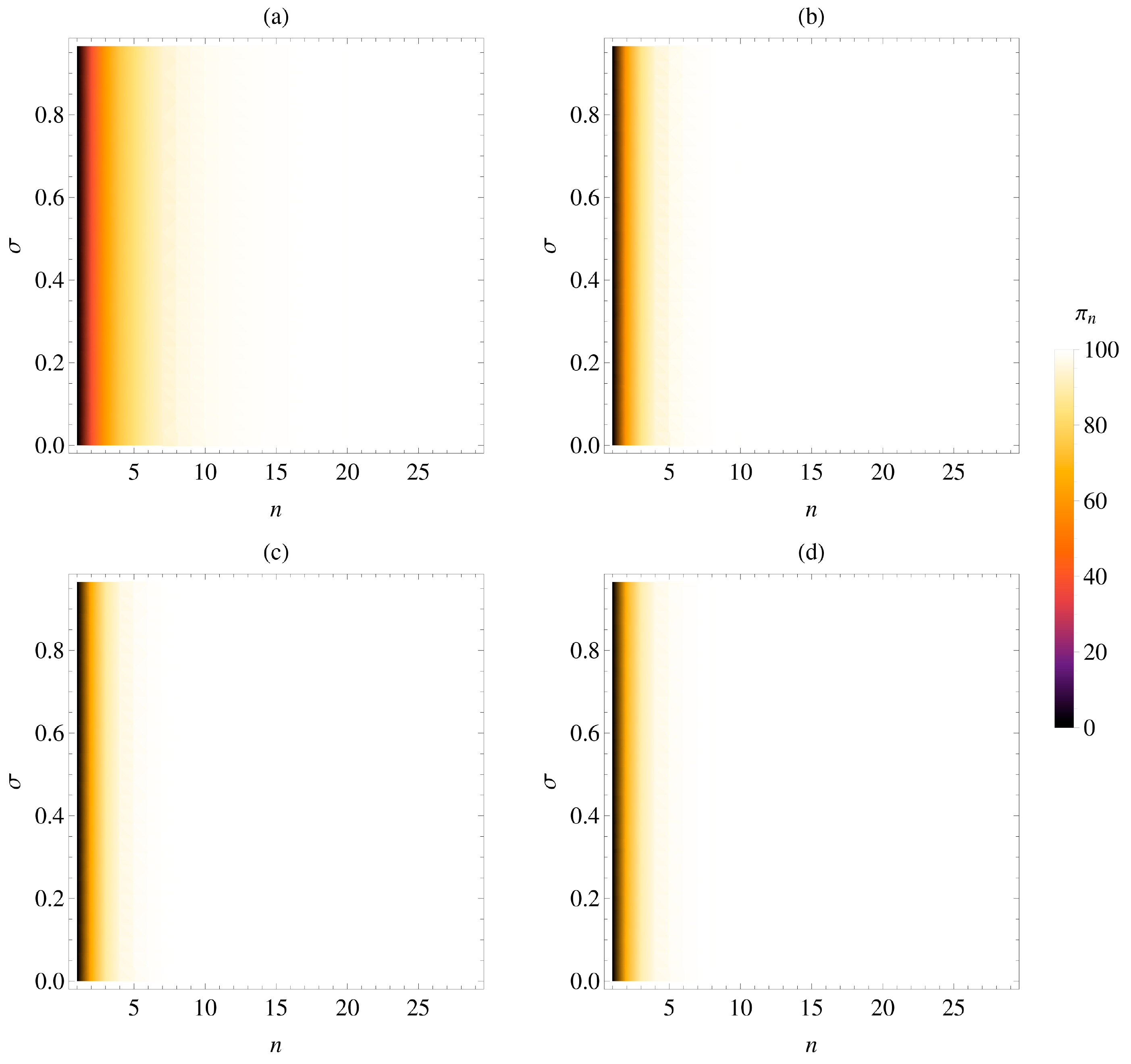}
	\caption{Value of $\pi _{n}$ with respect to $\Phi _{\mathbb{E}[Z^{k}]_{p}}(\lambda _{2},...,\lambda _{n},\sigma;p)$ and the disorder parameter $\sigma$ for (a) $p=1.5$, (b) $p=3.0$, (c) $p=4.0$, (d) $p=5.0$.}
	\label{figexpected3}
\end{figure}

As before, we compute the behavior of (\ref{percentage1111}) with respect to the disorder parameter $\sigma$. In FIG. \ref{figexpected3}, we have that the region where we can achieve the 40\%-60\% performance improvement with a low reasonable added links quantity has increased; even in the great disorder limit we have a good behavior for few added links. 
 
\begin{figure}[htb]
	\centering
	\includegraphics[width=.72\columnwidth]{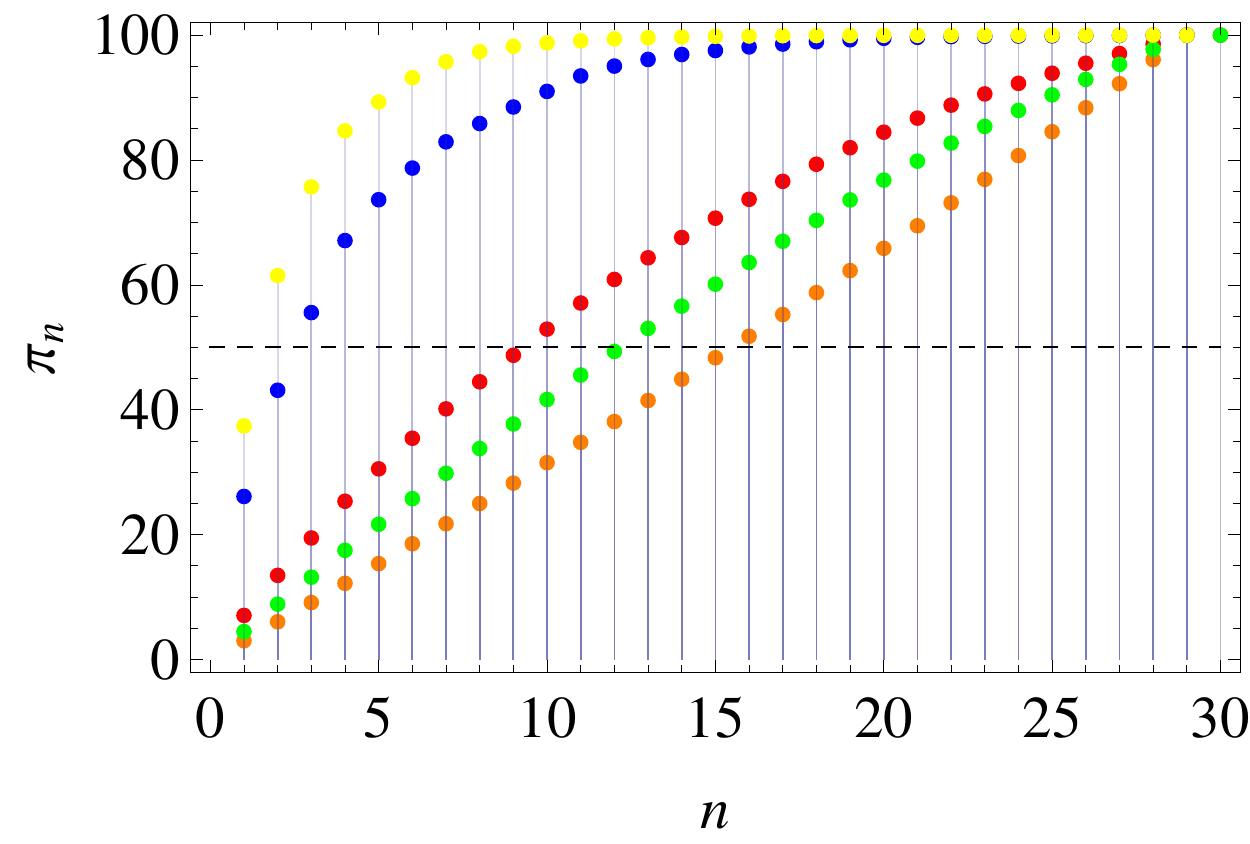}
	\caption{Comparison of different SMPs, expected value of the replica generalized partition function $\Phi _{\mathbb{E}[Z^{k}]_{p=1.5}}$ (yellow), partition function $\Phi _{Z}(p=0.1)$ (blue), spectral zeta function (\ref{spectralzetafunction1}) $\zeta _{1}$ (red), expected transient covariance (\ref{expectedtransientoutput}) $\tau _{t=0.1}$ (green), and $\Phi _{Z}(p=1.5)$ (orange).}
	\label{figcomparison1}
\end{figure}

A comparison of the advantage of our SPM with respect to other known SPMs are depicted in FIG. \ref{figcomparison1}. We can see that our set of parameters associated to our new set of SPM give us the possibility to obtain ranges and values that improve the behavior of our objects and show better results with respect to other spectral functions. We can see that we need to add 1 or 2 links to improve the network performance over the 50\%.

\begin{figure*}[htb]
	\includegraphics[width=0.9\textwidth]{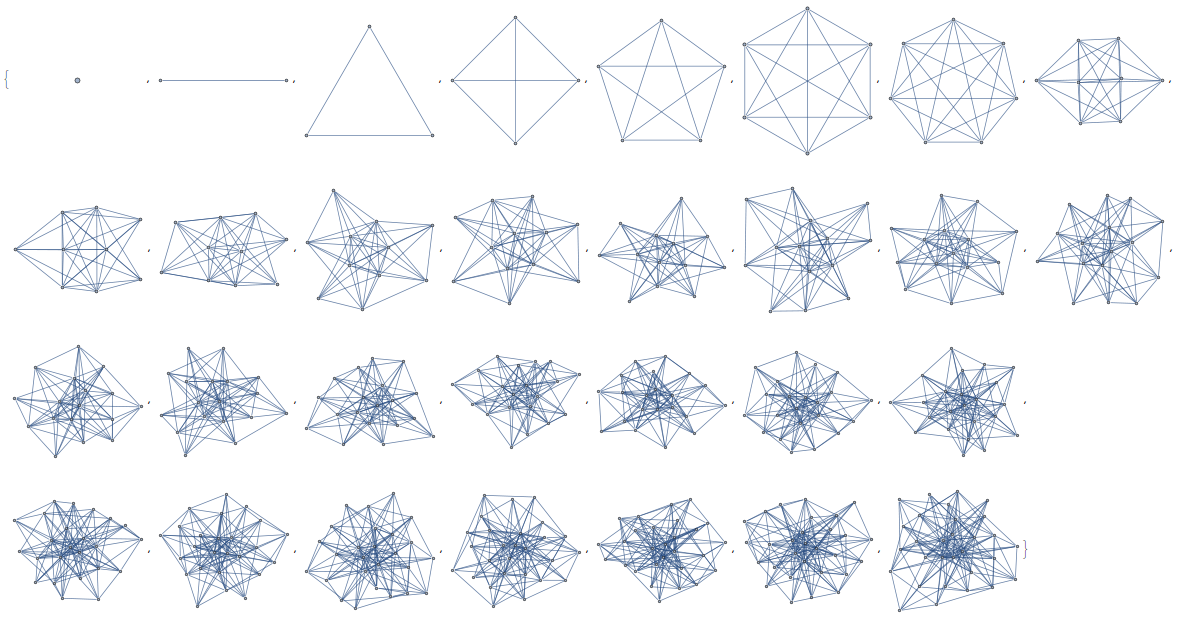}
	\caption{Subgraphs contribution to $\Phi _{F_{q}^{(p)}}$.}
	\label{netsub111}
\end{figure*}

\begin{figure}[htb]
	\includegraphics[width=0.9\columnwidth]{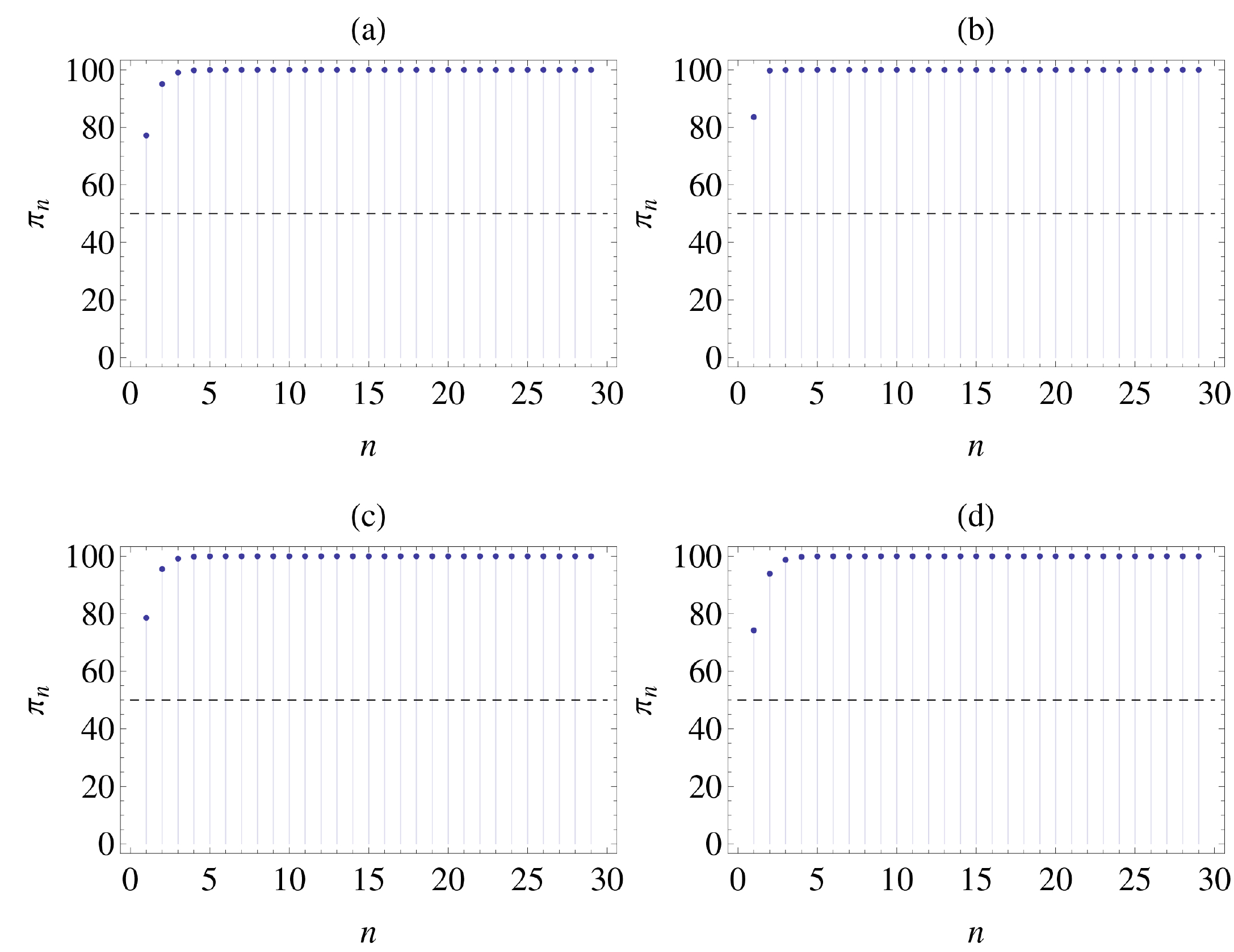}
	\caption{Value of $\pi _{n}$ with respect to $\Phi _{F_{q}^{(p)}}$ for (a) $\sigma =0.1$ $p=1.5$ and 1 field in the second phase, (b) $\sigma =0.1$ $p=1.5$ and 29 fields in the second phase, (c) $\sigma =0.1$ $p=1.5$ and 15 fields in the second phase, (d) $\sigma =0.5$ $p=1.5$ and 15 fields in the second phase.}
	\label{figenergy111}
\end{figure}

Finally, for the free energy of a graph with disorder, results show considerable improvement with respect the other studied SPMs. As we previously mentioned, each contribution of the series representation (\ref{energy333}) is interpreted as a subgraph of the total graph. For example, let us consider the network and its subsystems shown in FIG. \ref{netsub111}. Each step will contribute to the total free energy as it is shown. The improvement could be originated by consider the dynamics of each subgraph and the phase transition controlled by the new parameter $m_{0}^{2}$. In FIG. \ref{figenergy111} are depicted the results for different configurations of replica fields. We can evidence that these configurations improve the behavior of the aforementioned SPM. 

\begin{figure*}[htb]
\centering
  \begin{tabular}{@{}ccc@{}}
    \includegraphics[width=.25\textwidth]{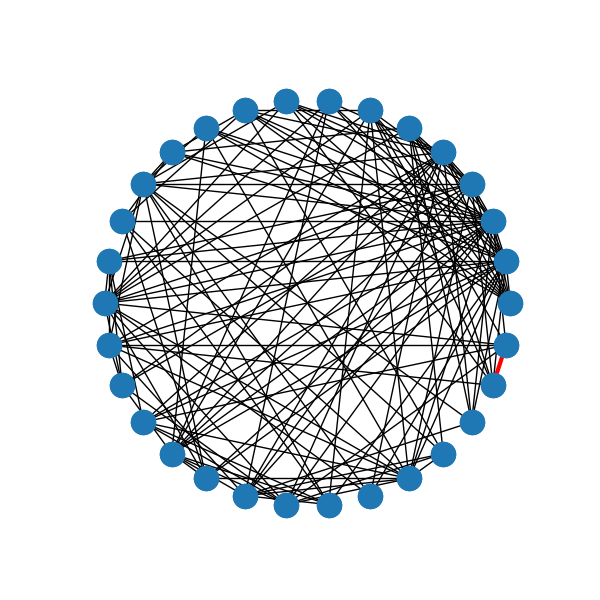} &
    \includegraphics[width=.25\textwidth]{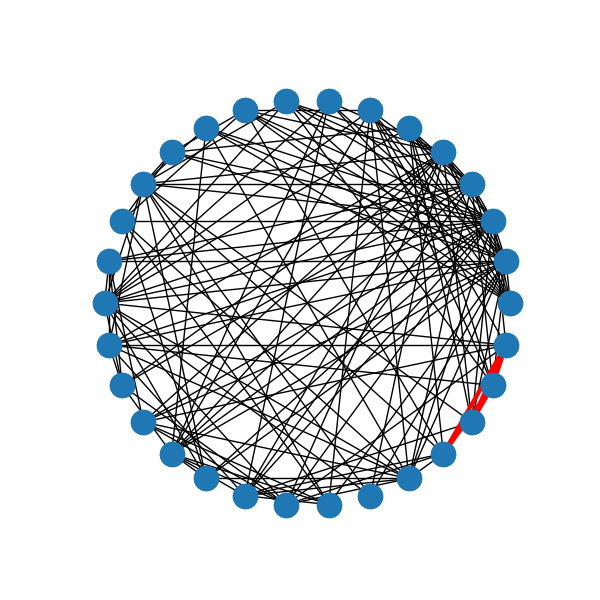} &
    \includegraphics[width=.25\textwidth]{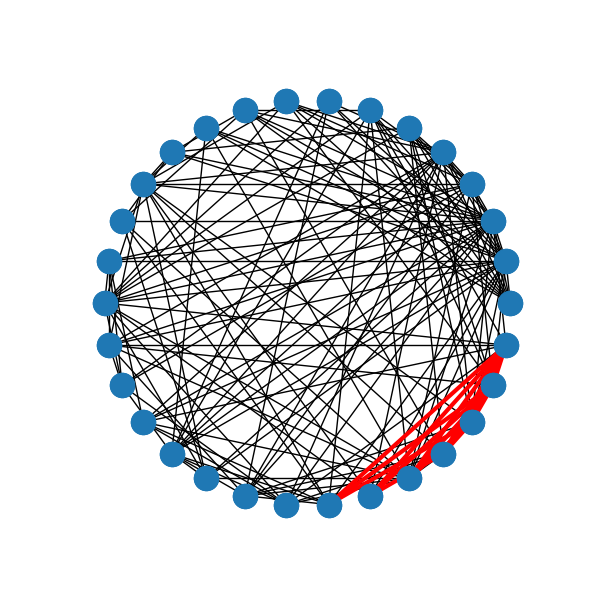}\\
    \includegraphics[width=.25\textwidth]{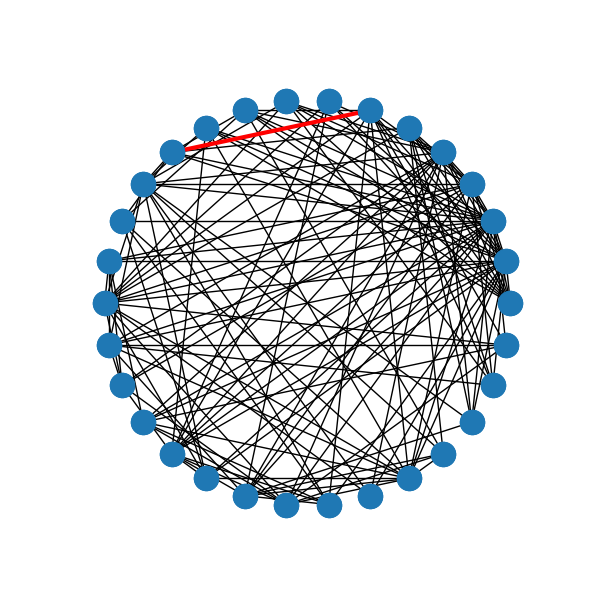} &
    \includegraphics[width=.25\textwidth]{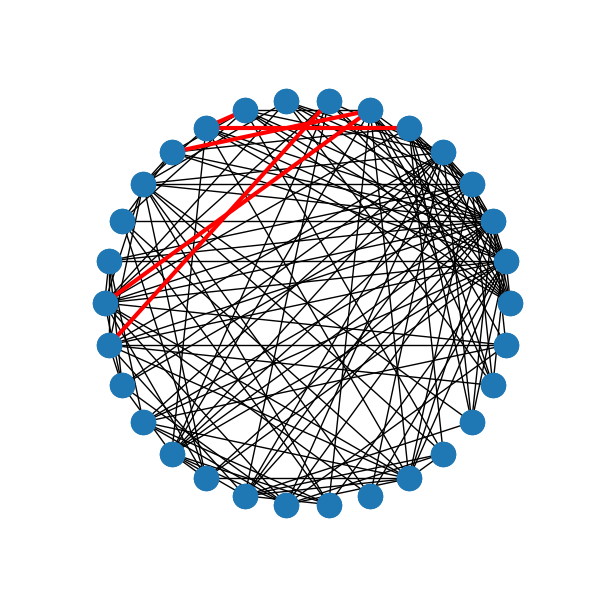} &
    \includegraphics[width=.25\textwidth]{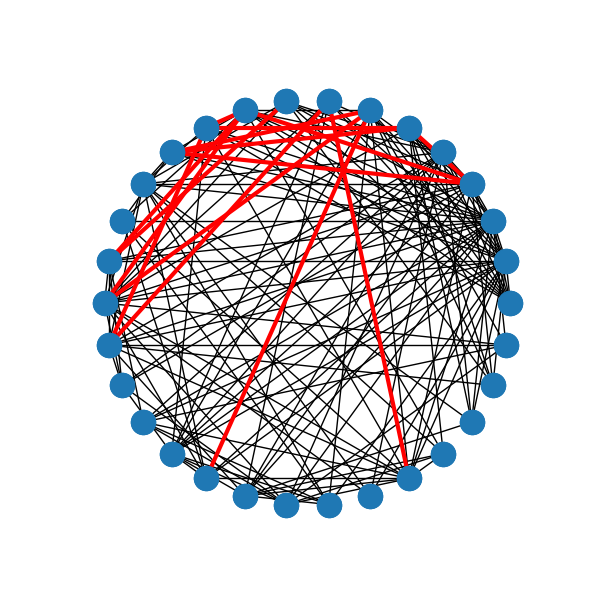} \\

  \end{tabular}
  \caption{Sensivity of location of optimal links as a function of SPM (\ref{phiz2}) with $p=0.1$ and $\sigma = 0.1$ (first row) vs. SPM (\ref{expect22}) with $p=1.5$ and $\sigma = 0.1$ (second row).}
  \label{fig4}
\end{figure*}

On the other hand, in Fig. \ref{fig4} we show the results of apply the growing Algorithm 1 and 2 to a Albert-Barab\'asi network with $k=30$. The set candidate links is the set of all possible links in the coupling graph, i.e., $|\mathcal{E}_{c}|=\frac{1}{2}k(k-1)$. All the candidate links have an identical weight. The first row is showing the pattern of growing when we use the SPM (\ref{phiz2}) with $p=0.1$ and $\sigma = 0.1$. In  this case, we have a pattern that profiles an \emph{ordered} growing promoting clusters generation. The case of SPM (\ref{expect22}) with $p=1.5$ and $\sigma = 0.1$ the ordered growing have disappear so far and this object tends to a homogeneous growing. In both cases the measures of coherence, captured by the $\mathcal{H}_{2}$-norm, and spectral zeta $\zeta _{q}$ increases up to 40\% with few links. The proper SPMs (\ref{phiz2}) and (\ref{expect22}) increases up to 50\% when we increase the weight value (see definitions in Appendix \ref{subappend111}).

\begin{figure*}[htb]
\centering
  \begin{tabular}{@{}ccc@{}}
    \includegraphics[width=.21\textwidth]{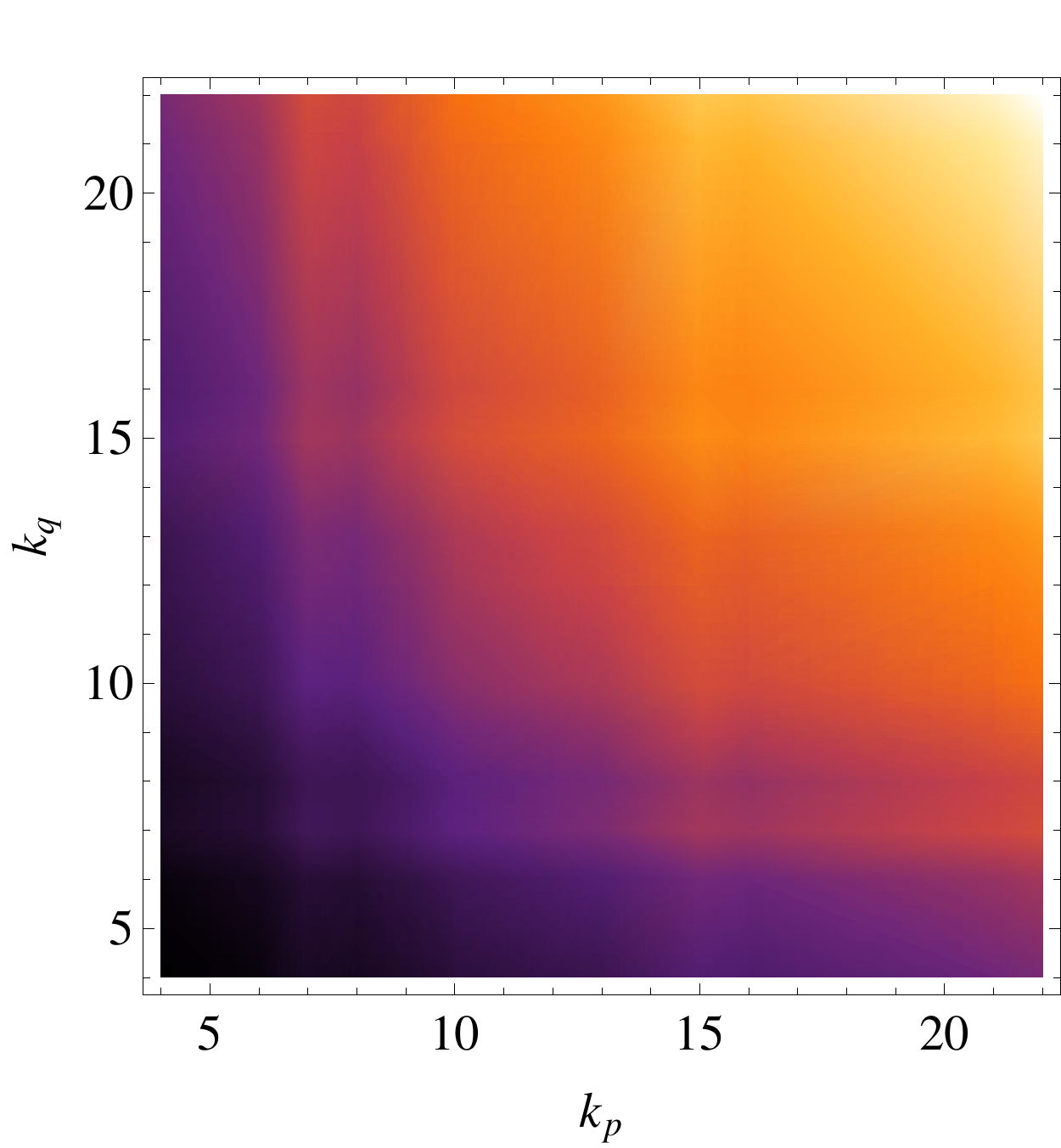} &
    \includegraphics[width=.21\textwidth]{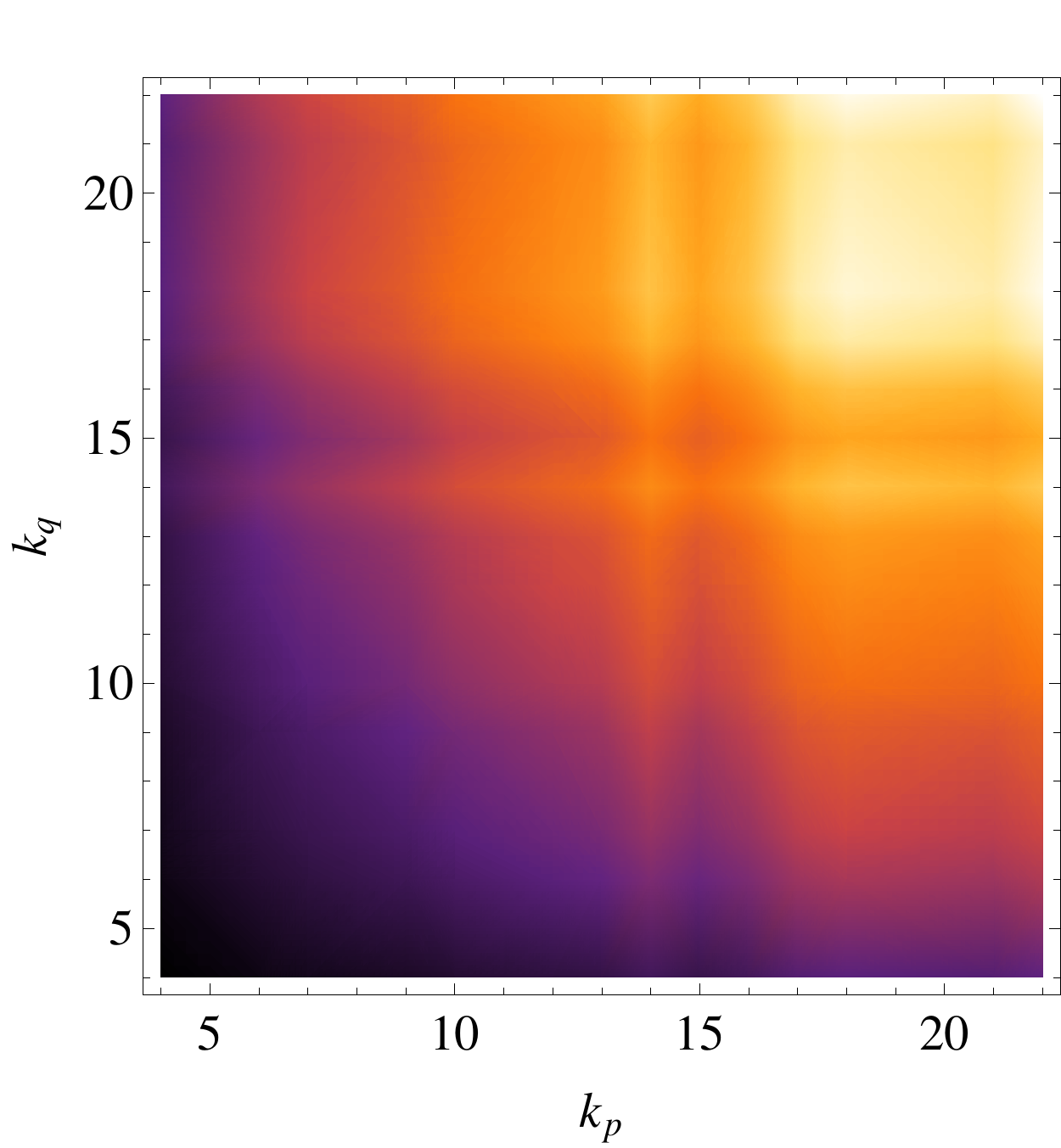} &
    \includegraphics[width=.21\textwidth]{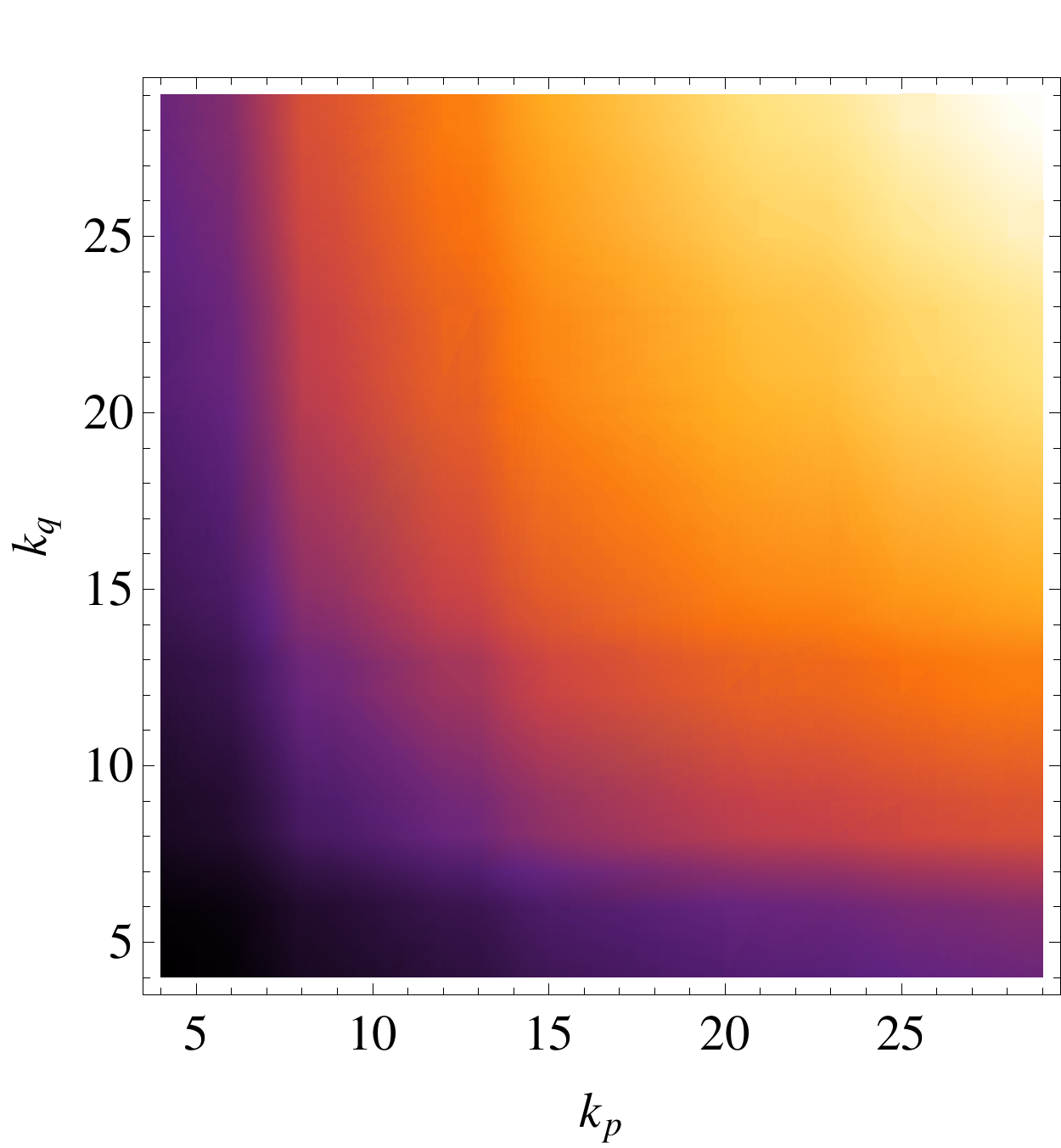} \\
    \includegraphics[width=.24\textwidth]{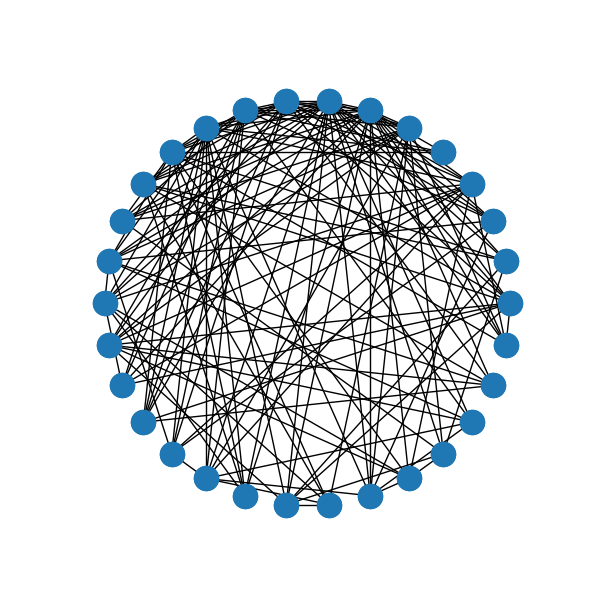} &
    \includegraphics[width=.24\textwidth]{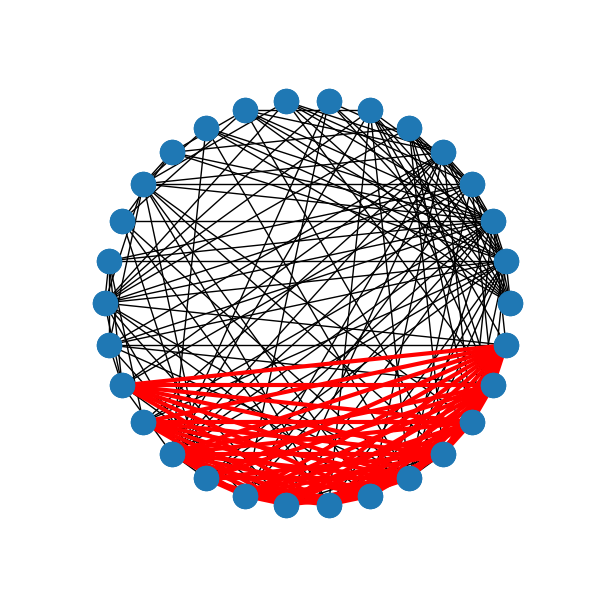} &
    \includegraphics[width=.24\textwidth]{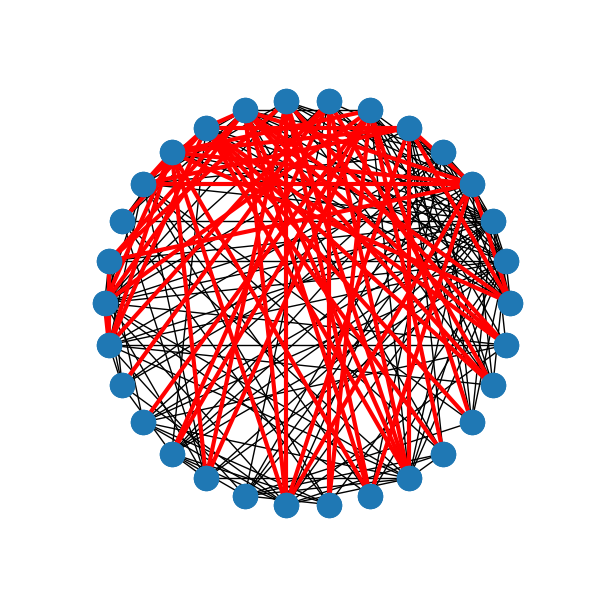} \\
	\multicolumn{3}{c}{\includegraphics[width=.21\textwidth]{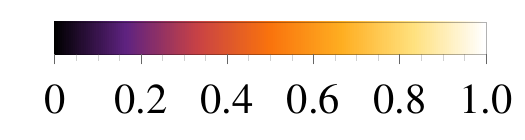}}
  \end{tabular}
  \caption{Communicability (\ref{communica1111}) of the original network (first column), grown network with SPM (\ref{phiz2}) with $p=0.1$ and $\sigma = 0.1$ (second column) and grown network with SPM (\ref{expect22}) with $p=1.5$ and $\sigma = 0.1$ (third column), for analysis of network desing problem. The first row shows the communicability respect to node degree. The final row of each column correspons to the original network and the resultant networks after adding 60 links with its respective SPM.}
  \label{fig5}
\end{figure*}

The communicability $G_{c}(p,q)$ between the nodes $p$ and $q$ in the network (associated with the Green function of a network)~\cite{estrada2} defined by
\begin{equation}
G_{c}(p,q)=\sum _{j=1}^{k}\psi _{j}(p)\psi _{j}(q)e^{\alpha _{j}}
\label{communica1111}
\end{equation}
where $\psi _{j}(p)$ is the $p$th element of the $j$th orthonormal eigenvector of the adjacency matrix associated with the eigenvalue $\alpha _{j}$, is depicted in FIG. \ref{fig5}. The communicability (\ref{communica1111}) communicability for degree nodes are depicted in the row 1. The original newtork is in the last row of this column. The second  column is for SPM (\ref{phiz2}) with $p=0.1$ and $\sigma = 0.1$ and the third column is for SPM (\ref{expect22}) with $p=1.5$ and $\sigma = 0.1$. The final row corresponds to the final result after adding 60 weighted links. Here is more easy to evidence the growing patterns induced by each SPM. For the three situations we have a typical pattern of assortative communicability as reported in~\cite{estrada2}. The assortative communicability may appear in homogeneous networks where the hubs can communicate to each other with or without structural bottlenecks~\cite{estrada23}; we can see that for the case of the final pattern induced by SPM (\ref{phiz2}) we have a hub formation and communication without structural bottlenecks, where can be manifest interhub communication by indirect routes~\cite{estrada24}, meanwhile for (\ref{expect22}), the communicabillity has grown in a homogeneous manner. This difference of growing dynamics shows the different scenarios where we could use the elements of our new set of SPM.

Finally, we study the communicability and network communities by with the quantity $\Delta G(p,q)$ defined by 
\begin{eqnarray}
\Delta G(p,q)&=&G(p,q)-\psi _{1}(p)\psi _{1}(q)e^{\alpha _{1}}
\nonumber \\
&=& \sum _{j=2}^{k}\psi _{j}^{+}(p)\psi _{j}^{+}(q)e^{\alpha _{j}}+\sum _{j=2}^{k}\psi _{j}^{-}(p)\psi _{j}^{-}(q)e^{\alpha _{j}}
\nonumber \\
&\, &+\sum _{j=2}^{k}\psi _{j}^{+}(p)\psi _{j}^{-}(q)e^{\alpha _{j}}
\label{commun33}
\end{eqnarray}
where $\psi _{j}^{+}$ are the eigenvectors components with positive sign and $\psi _{j}^{+}$ with negative ones. The Eq. (\ref{commun33}) can be rewritten in virtue of the first and second term represent the intracluster communicability and the last term represents the intercluster communicability~\cite{estrada2}. Therefore we have
\begin{eqnarray}
\Delta G(p,q) &=& \sum _{j=2}^{\text{intracluster}}\psi _{j}(p)\psi _{j}(q)e^{\alpha _{j}}
\nonumber \\
&\, & - \left| \sum _{j=2}^{\text{intercluster}}\psi _{j}(p)\psi _{j}(q)e^{\alpha _{j}}\right| .
\label{commun44}
\end{eqnarray}

\begin{figure*}[htb]
\centering
  \begin{tabular}{@{}ccc@{}}
    \includegraphics[width=.24\textwidth]{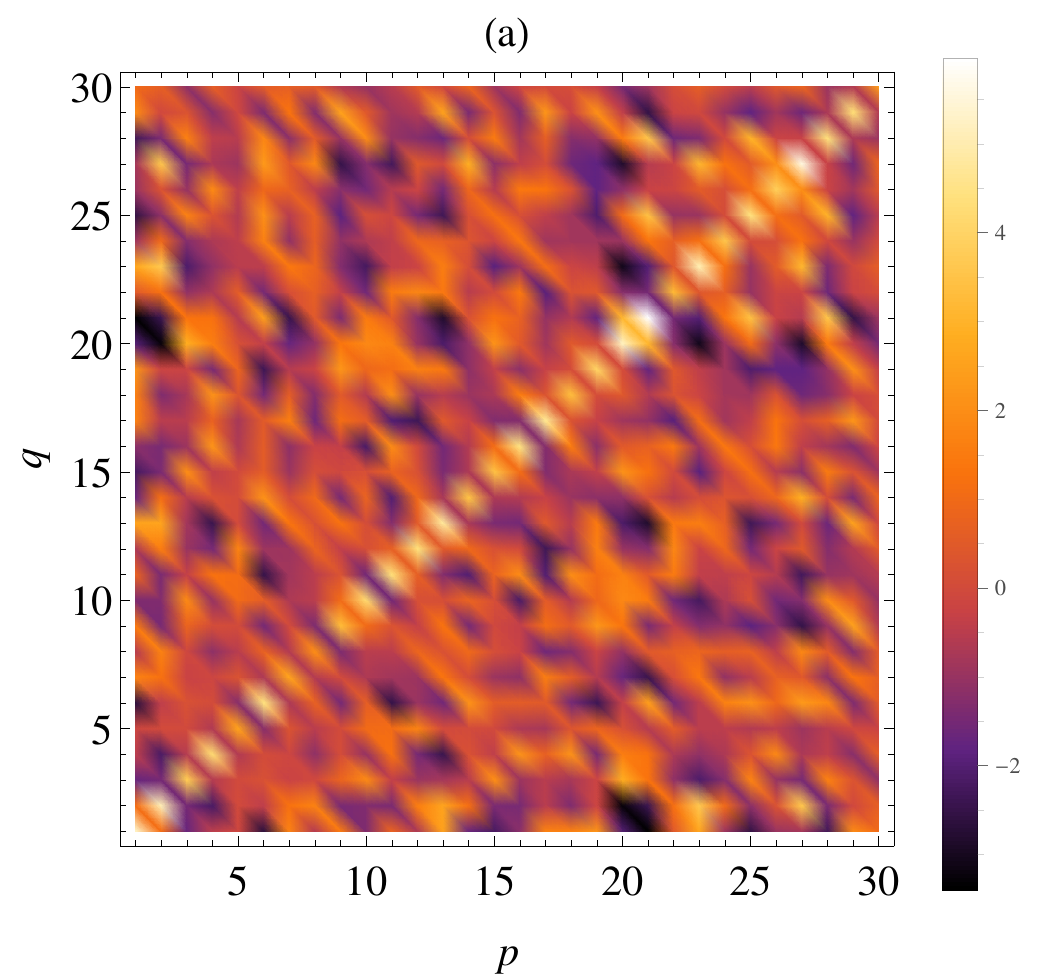} &
    \includegraphics[width=.24\textwidth]{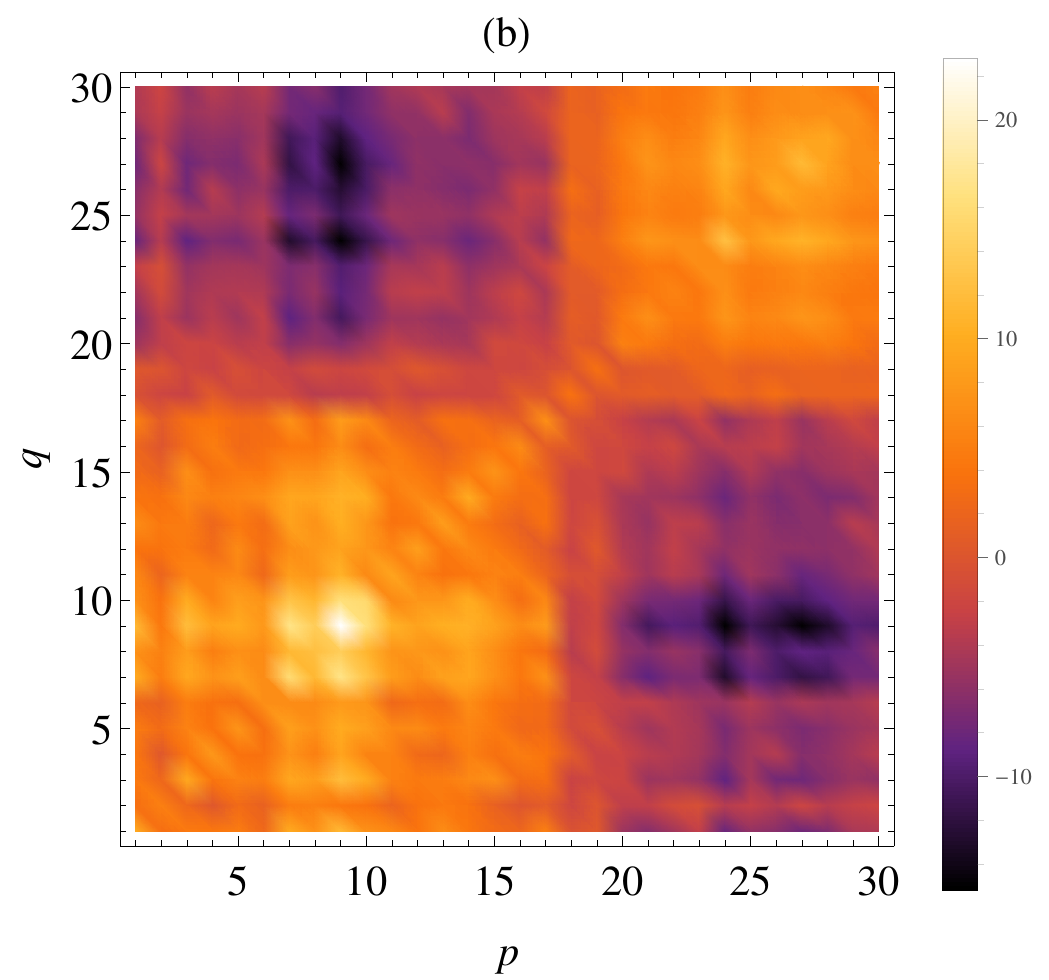} &
    \includegraphics[width=.24\textwidth]{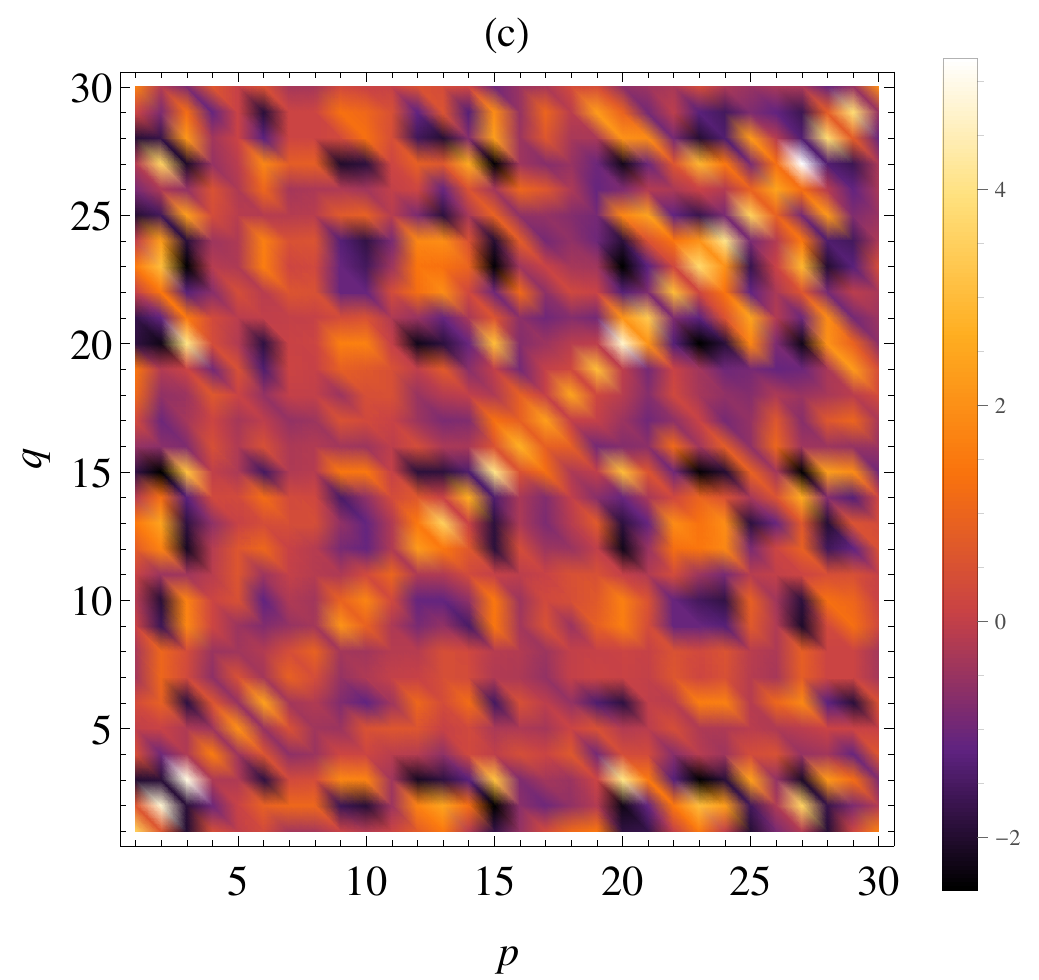} \\
  \end{tabular}
  \caption{Quantity $\Delta G(p,q)$ (\ref{commun44}) for  (a) the original network, (b) grown network with SPM (\ref{phiz2}) with $p=0.1$ and $\sigma = 0.1$, and (c) grown network with SPM (\ref{expect22}) with $p=1.5$ and $\sigma = 0.1$. We have a community formation for SPM (\ref{phiz2}).}
  \label{fig6}
\end{figure*}

In FIG. \ref{fig6} we show the quantity (\ref{commun44}) for the SPM (\ref{phiz2}) with $p=0.1$ and $\sigma = 0.1$ and SPM (\ref{expect22}) with $p=1.5$ and $\sigma = 0.1$. We have that we obtain a community formation when we use the SPM (\ref{phiz2})~\cite{estrada25}; we can evidence in FIG. \ref{fig6} (b) an usual pattern of community structre as reported in~\cite{estrada2}. Otherwise, is clear the lack of communities formation.

\section{Concluding remarks}
\label{conclusions}
In this paper, we consider a disorder field theory defined in a $d$-dimensional Euclidean space as a background to construct a new set of SPM to define proper measurements that may capture essential characteristics of a network. Furthermore, they can be used in the network design problem. It is known that when we have a disorder field interacting with scalar fields, the ground state configurations of fields are defined by a saddle-point equation, where the solutions of such an equation depend on particular configurations of disorder. Generally, a way to study these kind of systems is by averaging the free energy of system over the disorder field. The replica method has been used for this purpose. However, recently an alternative approach to compute this average called the Distributional Zeta Function (DZF) has been used with success for a several physical systems with disorder. In this approach, the leading contribution to the average free energy is expressed as a series of integer moments of the partition function of the model. Each term of this series depicts a replica field theory. Since all replicas contribute to average free energy, it is shown that under this formalism there exists a spontaneous symmetry breaking between the replica field theories. 

Using this framework, we have extended this approach to networks to construct a set of Laplacian spectral functions and evaluate the conditions that they must accomplish to be a SPM. We have constructed and studied four spectral functions, i.e., i) The generalized field network partition funcion; ii) The Distributional Zeta Function of a field network; iii) The Expected value of the replica field network partition function; and iv) The quenched Free Energy of a field network. We have taken advantage from the disordered-induced interaction, the functional form of DZF, the series representation of quenched free energy, and the spontaneous symmetry breaking mechanism, and its physical implications to generate a set of parameters that maximize the improvement action of each new SPM. Each object can be used to study different topological and thermodynamics issues of networks in an analogous form that have been used the energy, entropy, temperature, and density matrix before. 

We have shown that these DZF-based SMPs exhibit great advantages for obtaining a performance improvement with a few operations over the network. We obtained that after the process of adding links, the performance measurements of interest reach a significant percentage increase, and our objects can be used as a performance measurement to explore other network issues of interest. Furthermore, by studing the communicability of the resulting grown networks (associated with the Green function of network), we showed that each new object induce a different growing pattern that could be use for different design purposes. 

A crucial point is that, in the case of the quenched free energy of a field network, each subgraph (each replica system) with its own dynamics is contributing to the formation of SPM. It shows a great advantage with respect to other spectral functions. We showed that a high enhancement percentage can be obtained with a very few number of steps. This shows that the spontaneous symmetry breaking mechanism discovered under the DZF formalism can bring us a path to use different phase transitions situations to construct and evalue performance measures. Furthermore, we want to point out that the extension of network science to field theory could bring us a more general scenario to study issues that have been unnoticed by the current tools.

A natural continuation of this paper is to study other Euclidean models (such as Landau-Ginzburg model~\cite{ma1, sherr1, tarjus1}, Gross-Pitaevskii model~\cite{piit1, gross11, calzeta0}, Euclidean Schwarzschild and other curved manifolds~\cite{syman1, hartle1, gibbons1, calzeta1, calzeta2}) in order to obtain new elements for the set of DZF-based SPMs and extract out its advanges from its physical consequences. Another continuation is to evalue different phase transitions problems and entanglement networks of quantum systems of interets (such as qubits~\cite{biamonte1, elias1, potes1, garnerone1, perseguers1} or biological light-harvesting complexes~\cite{kolli1, stones1, curut1, qin1, plenio1}). Finally, the objects that we have constructed here can be used to study design, formation, growing, and robustness of real life networks to obtain a depper understanding of their complexity. We can relate these objects with an existent robustness techniques to unveil structural vulnerability and improve network resiliency against cascading collapse~\cite{eduardom1}. These issues are under investigation by the authors.

\section*{Acknowledgments}

This paper was partially supported by the \emph{VIII Convocatoria para el Desarrollo y Fortalecimiento de los Grupos de Investigaci\'on en Uniminuto} with code C119-173 and Industrial Engineering Program from the Corporaci\'on Universitaria Minuto de Dios (Uniminuto, Colombia).

\section*{Appendix}
In this appendix we include further developments and several theorems of interest that bring us tools to develop our DZF-based SPMs.
\appendix

\section{Hermiticity of normalized graph Laplacian operator}
\label{appendix1}

In order to study the properties of (\ref{laplacianmatrix111}) and its usefulness to establish a relation with the Hamiltonian operator, let us firstly define the inner product $(\cdot , \cdot)$ over a Hilbert space is as
\begin{equation}
(f_{n}(x),f_{m}(x))=\int _{a}^{b}p(x)f^{*}_{n}(x)f_{m}(x) .
\end{equation}

Furthermore, the orthonormality of weight $p(x)$, over an interval $[a,b]$, between two functions $f_{n}(x)$ and $f_{m}(x)$ is defined by
\begin{equation}
\int _{a}^{b}dx p(x)f^{*}_{n}(x)f_{m}(x) = \delta _{n m} .
\end{equation}

Therefore, the following theorem allows us to replace formally a hermitic operator by the normalized graph Laplacian assumed too as a hermitic operator.

\emph{Theorem A.1.:} Let $f_{\mu}(x)$ be a solution of the following eigenvalues problem, $\forall \mu \in \mathbb{N}$, 
\begin{equation}
\tilde{L}f_{\mu}(x)=\tilde{\lambda} _{\mu}f_{\mu}(x)
\end{equation} 
then $\tilde{L}$ is a hermitic operator, that is,
\begin{equation}
(f_{\mu}(x), \tilde{L}f_{\nu}(x))=(\tilde{L}f_{\mu}(x),f_{\nu}(x)) ,
\end{equation}

\emph{Proof:} To show that $\tilde{L}$ is hermitic, we have to show
\begin{equation}
\int _{a}^{b} dxf_{\nu}^{T}(x)\tilde{L}f_{\mu}(x)=\int _{a}^{b} dx [\tilde{L}f_{\nu}(x)]^{T}f_{\mu}(x) .
\end{equation}

The matrix elements are
\begin{equation}
[\tilde{L}f_{\mu}(x)]_{i}=\sum _{j} \tilde{l}_{ij}(f_{\mu}(x))_{j} .
\end{equation}

Then, the scalar product yields
\begin{equation*}
f_{\nu}^{T}(x)\tilde{L}f_{\mu}(x)=\sum _{ij}\tilde{l}_{ij}(f_{\nu}(x))_{i}(f_{\mu}(x))_{j} .
\end{equation*}

Analogously
\begin{equation*}
[\tilde{L}f_{\nu}(x)]^{T}f_{\mu}(x)=\sum _{ij}\tilde{l}_{ij}(f_{\mu}(x))_{i}(f_{\nu}(x))_{j} .
\end{equation*}

Thus, we have to show that
\begin{equation}
\sum _{ij}\tilde{l}_{ij}(f_{\mu}(x))_{j}(f_{\nu}(x))_{i}=\sum _{ij}\tilde{l}_{ij}(f_{\mu}(x))_{i}(f_{\nu}(x))_{j}.
\label{eq11}
\end{equation}

From the matrix elements (\ref{tildee}) and the expression (\ref{eq11}), we have that, for $i=j$,
\begin{equation}
\sum _{i}(f_{\mu}(x))_{i}(f_{\nu}(x))_{i}=\sum _{i}(f_{\mu}(x))_{i}(f_{\nu}(x))_{i} .
\end{equation}

For $i\neq j$, since the symmetry and the mude indices, we have
\begin{eqnarray*}
\sum _{ij}\tilde{l}_{ij}(f_{\mu}(x))_{j}(f_{\nu}(x))_{i}&=&\, \sum _{ij}\tilde{l}_{ji}(f_{\mu}(x))_{j}(f_{\nu}(x))_{i}
\nonumber \\
&=&\, \sum _{ij}\tilde{l}_{ij}(f_{\mu}(x))_{j}(f_{\nu}(x))_{i} .
\end{eqnarray*}

Then,
\begin{eqnarray}
0&=&\, (-\tilde{\lambda} _{\mu}+\tilde{\lambda} _{\nu})\int _{a}^{b}dx p(x) f_{\nu}^{T}(x)f_{\mu}
\nonumber \\
&=&\, (f_{\mu}(x),\tilde{L}f_{\nu}(x)) - (\tilde{L}f_{\mu}(x),f_{\nu}(x)) .
\end{eqnarray}

Therefore the graph Laplacian operator is hermitic $\blacksquare$.

\emph{Corollary A.2.:} The functions that accomplish $\tilde{L}f_{\mu}(x)=\tilde{\lambda} _{\mu}f_{\mu}(x)$ generate a basis in a Hilbert space.
\section{Theorems for determine conditions to be a SPM}
\label{appendix2}
In this Appendix we list the important theorems presented in~\cite{siami5} to explore the conditions that must accomplish a candidate function to be a SPM.

\emph{Theorem A.3.:} Let $L\in \mathcal{L}_{k}$. Assume that $\varphi :\mathbb{R}_{+}\rightarrow \mathbb{R}$ is a decreasing convex function. Then, the following spectral function:
\begin{equation}
\rho (L)=\sum _{i=2}^{k}\varphi (\lambda _{i})
\end{equation}
is a SPM. In addition, if $\varphi$ is also a homogeneous function of order $-\kappa$ with $\kappa > 1$, then the following spectral function:
\begin{equation}
\rho (L)=\left[ \sum _{i=2}^{k}\varphi (\lambda _{i}) \right] ^{\frac{1}{\kappa}}
\end{equation}
is also a SPM.

It is known that the Laplacian eigenvalues of a network are characterized by the global features of the intrinsic coupling graph. This is the reason why every performance measure that satisfies the aforementioned definition is labeled with adjective \emph{systemic}.

There is a theorem that compute the theoretical bounds for the best achievable values for the performance measure. Denoting the optimal cost value by $r_{k}^{*}(\varpi)$, the theorem reads

\emph{Theorem A.4.:} Suppose that an ordered set of Laplacian eigenvalues $\lambda _{2}\leq \cdots \leq \lambda _{k}$ is given. Let $\mathcal{E}_{c}$ be a set of candidate links endowed with a weight function $\varpi : \mathcal{E}_{c}\rightarrow \mathbb{R}_{+}$. Consider a design parameter $1\leq n\leq k-1$. Therefore the following inequality 
\begin{equation*}
r_{n}^{*}(\varpi) > \Phi (\lambda _{n+2},...,\lambda _{k},\infty,...,\infty)
\end{equation*}
holds for all weight functions $\varpi$. For $k\geq n$, all lower bounds are equal to $\Phi (\infty,...,\infty)$. Furthermore, if the SPM can be expressed in the following decomposable form
\begin{equation*}
\rho (L)=\sum _{i=2}^{k}\varphi (\lambda _{i})
\end{equation*} 
being $\varphi :\mathbb{R}_{+}\rightarrow \mathbb{R}$ a decreasing convex function with $\lim _{\lambda \rightarrow \infty}\varphi (\lambda)=0$, then the best achievable performance measure is characterized by
\begin{equation*}
r_{k}^{*}(\varpi) > \sum _{i=n+2}^{k}\varphi (\lambda _{i}) .
\end{equation*}

The value of this lower bound is given by Eq. (\ref{percentage1111}).

\subsection{Growing Algorithms}
\label{subappend111}
In~\cite{siami5} the authors use SPM to growing networks by the combinatorial optimization problem of minimize $\rho (L+\hat{L})$ subject to a definition of a set of all possible appended subgraphs, being $\hat{L}$ a Laplacian matrix of this set. The resulting network with Laplacian matrix $L+\hat{L}$ is referred to as the augmented network. We have that the candidate link set $\mathcal{E}_{C}$ contains information about the authorized locations to establish new feedback interconnections in the network. We have two kinds of growing algorithms based on linearization and greedy approximation. 

\textbf{Algorithm 1:} For the first algorithm, based on linearization, we have to set the Laplacian matrix of the original network $L$, a set of candidate links $\mathcal{E}_{C}$, a weight function of these links $\varpi$, and a design parameter $n$. For each cycle $\tau =1$ to $n$ we must find a link $e=\{ i,j\}\in \mathcal{E}_{C}$ that returns the maximum value for 
\begin{equation*}
\varpi(e)(\nabla _{L} \rho (L)_{ii}+\nabla _{L} \rho (L)_{jj}-\nabla _{L} \rho (L)_{ij}-\nabla _{L} \rho (L)_{ji}) ,
\end{equation*}
then, we set the solution $e^{*}$ to update our appended Laplacian matrix $\hat{L}$ as $\hat{L}=\hat{L}+\varpi (e^{*})L_{e^{*}}$, at the same time we update the set $\mathcal{E}_{C}$ as $\mathcal{E}_{C} = \mathcal{E}_{C}-\{ e^{*} \}$. The same procedure is performed for the next step.

\textbf{Algorithm 2:} For the other algorithm we set the Laplacian matrix of the original network $L$, a set of candidate links $\mathcal{E}_{C}$, a weight function of these links $\varpi$, and a design parameter $n$. For each cycle $\tau =1$ to $n$ we must find a link $e=\{ i,j\}\in \mathcal{E}_{C}$ that returns the maximum value for 
\begin{equation*}
\rho (\hat{L})-\rho (\hat{L}+\varpi (e)L_{e}) ,
\end{equation*}
then, we set the solution $e^{*}$ to update our appended Laplacian matrix $\hat{L}$ as $\hat{L}=\hat{L}+\varpi (e^{*})L_{e^{*}}$, at the same time we update the set $\mathcal{E}_{C}$ as $\mathcal{E}_{C} = \mathcal{E}_{C}-\{ e^{*} \}$. The same procedure is performed for the next step.

\section{Ramsey-based theorems for the eigenvalues of a HSN matrix}
\label{appendix3}

In this Appendix, we present a list of Ramsey-based theorems to evaluate the number of nonpositive eigenvalues of an HSN matrix and find relations that must accomplish the eigenvalues of (\ref{gamma1111}) in order to study when (\ref{expec12}) and (\ref{expect22}) are SPM. The complete discussions and demonstrations can be found in~\cite{charles1}-\cite{johnson1}.

\emph{Theorem A.5.:} Assume $A$ is a Hermitian matrix, and denote $B$ as the principal submatrix of $A$. Denote the eigenvalues of $A$ and $B$ by $\{ a _{i}\}$ and $\{ b_{i}\}$, respectively. Suppose that they have been arranged in nonincreasing order $a _{1}\geq \cdots \geq a _{n}$ and $b_{1}\geq \cdots \geq b_{n-1}$. Then
\begin{equation*}
a _{i}\geq b_{i} \geq a _{i+1},\quad \text{for}\quad i=1,2,...,n-1.
\end{equation*} 

Within the above theorem it can be concluded that if a matrix has a principal submatrix with $k$ nonpositive eigenvalues, thus the matrix itself has at least $k$ nonpositive eigenvalues. This will be especially helpful in the situation where $G$ is a complete graph or an empty graph, two kinds of graphs that naturally arise in Ramsey theory. Supposing that we wish to color the edges of a complete graph $G$ with $n$ colors, we may define the \emph{generalized Ramsey number} $R(r_{1},r_{2},...,r_{n})$ as the minimum number of vertices of the complete graph $G$ such that for some $i\in \{ 1,2,...,n \}$, there is an induced complete subgraph on $r_{i}$ vertices with all edges of color $i$. The existence of such number is assured by Ramsey's theorem.

With the idea of the generalized Ramsey number we can relate the off-diagonal entries of a $n$-by-$n$ HSM that are drawn from a fixed finite set with the number of nonpositive eigenvalues.

\emph{Theorem A.6.:} Denote $S$ as a finite set of nonnegative numbers. Assume that $k$ is a fixed positive integer. Then there is an $n$ such that all HSN matrices of order at least $n$ and with off-diagonal entries from $S$ have at least $k$ nonpositive eigenvalues.

\emph{Corollary A.7.:} Denote $S$ as a finite set of positive numbers. Assume that $k$ is a fixed positive integer. Then, there is an $n$ such that all HSN matrices of order at least $n$, and with off-diagonal entries from $S$, have at least $k$ negative eigenvalues.

The above corollary can be generalized to infinite sets with a mild, but improved, restriction.

\emph{Theorem A.8.:} Let $0< \epsilon < 1$ be a real number. Assume that $k$ is a fixed positive integer. Then, there is an $n$ such that all HSN matrices of order at least $n$, and with off-diagonal entries from $(\epsilon ,1]$, have at least $k$ negative eigenvalues.

\emph{Corollary A.9.:} Let $0< \epsilon < 1$ be a real number. Assume that $k$ is a fixed positive integer. Then, there is an $n$ such that all HSN matrices of order at least $n$, and with off-diagonal entries from $\{ 0 \} \bigcup (\epsilon ,1]$ have at least $k$ negative eigenvalues.

Finally we have a result showing that more sophisticated division gives us a much better bound for $\epsilon$.

\emph{Theorem A.10:} Consider $n$ and $2\leq k \leq n-1$ two positive integers. Denote $c$ as the smallest integer for which
\begin{equation*}
n\leq R(k+1,k+1,...,k+1) .
\end{equation*}

Setting $\epsilon = [k/(k+1)]^{c}$, we shall have that all HSN matrices of order at least $n$, and with off-diagonal entries from $(\epsilon ,1]$, have at least $k$ negative eigenvalues.

For the first part of (\ref{expect14}) we can evidence that $(3\sigma ^{2}\Phi _{\mathbb{E}[Z^{k}[\lambda _{i}(k)]]}/4) v^{T}\Lambda v \geq 0$, thus the following inequality holds
\begin{equation}
v^{T}\nabla _{\lambda}^{2}\Phi _{\mathbb{E}[Z^{k}[\lambda _{i}(k)]]} v \geq \frac{\sigma ^{2}}{4}\Phi _{\mathbb{E}[Z^{k}[\lambda _{i}(k)]]}v^{T}\Gamma v .
\end{equation}

We shall study the nature of $v^{T}\Gamma v$ with the aforementioned theorems. Thus, there exists a fixed $n=k-1$ integer such that $n\leq R(k,k,...,k)$ that holds the quantity of negative eigenvalues of our $k$-by-$k$ HSN matrix (\ref{gamma1111}) in exactly $k-1$. Then the matrix (\ref{gamma1111}) is indefinite.

Knowing the nature of its eigenvalues, we can stablish a set of cases for our study. Let the eigenvalues of (\ref{gamma1111}) be denoted by $\{ \gamma _{i} \}$ and assume that they have been arranged in nonincreasing order $\gamma _{i}\geq \cdots \geq \gamma _{k}$. Then if $v^{T}\Gamma v \leq 0$ the eigenvalues must accomplish the following relation
\begin{equation}
0 \leq \gamma _{1}v_{1}^{2} \leq \sum _{i=2}^{k}\gamma _{i}v_{i}^{2} .
\end{equation}

Otherwise, if $v^{T}\Gamma v \geq 0$, we have that
\begin{equation}
0 \leq \sum _{i=2}^{k}\gamma _{i}v_{i}^{2} \leq \gamma _{1}v_{1}^{2} .
\end{equation}




\begin{thebibliography}{99}
\bibitem{wu1}  Wu, M. Barahona, Y. J. Tan, and H. Z. Deng. \href{http://dx.doi.org/10.1109/TSMCA.2011.2116117}{IEEE Trans. Syst., Man, Cybern. A \textbf{41}, 1244 (2011).}
\bibitem{abbas1} W. Abbas and M. Egerstedt, \href{https://www.sciencedirect.com/science/article/pii/S147466701534814X?via=ihub}{IFAC Proceedings Volumes \textbf{45}, 85 (2012).}
\bibitem{siami1} M. Siami and N. Motee, \href{10.1109/CDC.2013.6759860}{52nd IEEE Conference on Decision and Control 67 (2013)}.
\bibitem{wang1} X. Wang, E. Pournaras, R. E. Kooij, and P. Van Mieghem, \href{http://dx.doi.org/10.1140/epjb/e2014-50276-0}{Eur. Phys. J. B \textbf{87}, (2014).}
\bibitem{ye1} B. Ye, J. J. Jia, K. W. Zuo, and X. P. Ma, \href{http://dx.doi.org/10.1142/S0129183115500400}{Int. J. Mod. Phys. C \textbf{26}, 1550040 (2015)}.
\bibitem{duan1} B. Duan, J. Liu, M. Zhou, and L. Ma, \href{https://www.sciencedirect.com/science/article/abs/pii/S0378437115010730}{Phys. A \textbf{448}, 144 (2016).}
\bibitem{liang1} M. Liang, F. Liu, C. Gao, and Z. Zhang, \href{http://dx.doi.org/10.1109/DDCLS.2017.8068147}{6th Data Driven Control and Learning Systems (DDCLS)  23, 638 (2017).}
\bibitem{pizzuti1} C. Pizzuti and A. Socievole, \href{http://dx.doi.org/10.1007/978-3-030-05411-3_64}{Stud. Comput. Intell \textbf{812}, 807 (2019).}
\bibitem{wang2} S. Wang and J. Liu,  \href{http://dx.doi.org/10.1109/JSYST.2018.2835642}{IEEE Systems Journal \textbf{13}, 582 (2019).}
\bibitem{shang1} Y. Shang, \href{http://dx.doi.org/10.1109/TSMC.2017.2733545}{IEEE Trans. Syst. Man Cybern, Syst. \textbf{49}, 821 (2019).}
\bibitem{wang3} S. Wang and J. Liu, \href{https://www.sciencedirect.com/science/article/pii/S0020025518308922}{Inform. Sci. \textbf{478}, (2019).}
\bibitem{lu1} Y. Lu, Y. Zhao, F. Sun, and R. Liang, \href{http://dx.doi.org/10.1109/LCOMM.2019.2941940}{IEEE Commun. Lett. \textbf{23}, 2168 (2019).}
\bibitem{pizzuti2} C. Pizzuti, A. Socievole, and P. Van Mieghem, \href{http://dx.doi.org/10.1007/978-3-030-36687-2_61}{Stud. Comput. Intell. \textbf{881}, 735 (2020).}
\bibitem{girvan1} M. Girvan and M. E. J. Newman, \href{http://dx.doi.org/10.1073/pnas.122653799}{Proc. Nat. Acad. Sci \textbf{99}, 7821 (2002).}
\bibitem{mason1} O. Mason and M. Verwoerd, \href{http://dx.doi.org/10.1049/iet-syb:20060038}{IET Syst. Biol. \textbf{1}, 89 (2007).}
\bibitem{zhang1} H. Zhang, E. Fata, and S. Sundaram, \href{http://dx.doi.org/10.1109/TCNS.2015.2413551}{IEEE Trans. Control Netw. Syst. \textbf{2}, 310 (2015).}
\bibitem{bamieh1} B. Bamieh, M. R. Jovanovic, P. Mitra, and S. Patterson, \href{http://dx.doi.org/10.1109/TAC.2012.2202052}{IEEE Trans. Automat. Contr. \textbf{57}, 2235 (2012).}
\bibitem{young1} G. F. Young, L. Scardovi, and N. E. Leonard, \href{http://dx.doi.org/10.1109/CDC.2011.6161270}{50th IEEE Conference on Decision and Control and European Control Conference 1000 (2011).}
\bibitem{siami2} M. Siami and N. Motee, \href{https://www.sciencedirect.com/science/article/pii/S1474667015348497}{IFAC Proceedings Volumes \textbf{45}, 294 (2012).}
\bibitem{zelazo1} D. Zelazo, S. Schuler, and F. Allgower, \href{https://www.sciencedirect.com/science/article/pii/S0167691112002113}{Systems Control Lett. \textbf{62}, 85 (2013).}
\bibitem{ofa1} R. Olfati-Saber, J. A. Fax, and R. M. Murray, \href{http://dx.doi.org/10.1109/JPROC.2006.887293}{Proc. IEEE \textbf{95}, 215 (2007).}
\bibitem{siami3} M. Siami and N. Motee, \href{http://dx.doi.org/10.1109/ACC.2014.6859345}{American Control Conference, 5198 (2014).}
\bibitem{siami4} M. Siami and N. Motee, \href{http://dx.doi.org/10.1109/CDC.2014.7040189}{53rd IEEE Conference on Decision and Control, 5119 (2014).}
\bibitem{siami5} M. Siami and N. Motee, \href{http://dx.doi.org/10.1109/TAC.2017.2764447}{IEEE Trans. Automat. Contr. \textbf{63}, 2091 (2018).}
\bibitem{minello0} G. Minello, A. Torsello, and E. R. Hancock, \href{}{J. Complex Netw. \textbf{8}, (2020).}
\bibitem{barabasi1} R. Albert and A. L. Barabási, \href{http://dx.doi.org/10.1103/RevModPhys.74.47}{Rev. Mod. Phys. \textbf{74}, 47 (2002).}
\bibitem{bianconi0} G. Bianconi and A. L. Barabási, \href{http://dx.doi.org/10.1103/PhysRevLett.86.5632}{Phys. Rev. Lett. \textbf{86}, 5632 (2001).}
\bibitem{estrada1} E. Estrada and N. Hatano, \href{https://www.sciencedirect.com/science/article/abs/pii/S0009261407004058}{Chem. Phys. Lett. \textbf{439}, 247 (2007).}
\bibitem{park1} J. Park and M. E. J. Newman, \href{http://dx.doi.org/10.1103/PhysRevE.70.066117}{Phys. Rev. E \textbf{70}, 066117 (2004).}
\bibitem{javarone1} M. Alberto Javarone and G. Armano, \href{http://dx.doi.org/10.1088/1742-5468/2013/04/P04019}{J. Stat. Mech. \textbf{2013}, P04019 (2013).}
\bibitem{bianconi1} G. Bianconi, \href{http://dx.doi.org/10.1103/PhysRevE.87.062806}{Phys. Rev. E \textbf{87}, 062806 (2013).}
\bibitem{ostilli1} M. Ostilli and G. Bianconi, \href{http://dx.doi.org/10.1103/PhysRevE.91.042136}{Phys. Rev. E \textbf{91}, 042136 (2015).}
\bibitem{bianconi2} G. Bianconi, \href{http://dx.doi.org/10.1103/PhysRevE.96.012302}{Phys. Rev. E \textbf{96}, 012302 (2017).}
\bibitem{escolano1} F. Escolano, E. R. Hancock, and M. A. Lozano, \href{http://dx.doi.org/10.1103/PhysRevE.85.036206}{Phys. Rev. E \textbf{85}, 036206 (2012).}
\bibitem{ye11} C. Ye, C. H. Comin, T. K. DM. Peron, F. N. Silva, F. A. Rodrigues, L. da F. Costa, A. Torsello, and E. R. Hancock, \href{http://dx.doi.org/10.1103/PhysRevE.92.032810}{Phys. Rev. E \textbf{92}, 032810 (2015).}
\bibitem{ye2} C. Ye, A. Torsello, R. C. Wilson, and E. R. Hancock, \href{http://dx.doi.org/10.1007/978-3-319-18224-7_31}{Lecture Notes in Comput. Sci. \textbf{9069}, 315 (2015).}
\bibitem{minello1} G. Minello, A. Torsello, and E. R. Hancock, \href{http://dx.doi.org/10.1007/978-3-319-49055-7_5}{Lecture Notes in Comput. Sci. \textbf{10029}, 49 (2016).}
\bibitem{minello2} G. Minello, A. Torsello, and E. R. Hancock, \href{http://dx.doi.org/10.1109/ICPR.2016.7899855}{23rd International Conference on Pattern Recognition (ICPR) 1536 (2016).}
\bibitem{wang4} J. Wang, R. C. Wilson, and E. R. Hancock, \href{http://dx.doi.org/10.1007/978-3-319-49055-7_14}{Lecture Notes in Comput. Sci. \textbf{10029}, 153 (2016).}
\bibitem{wang5} J. Wang, R. C. Wilson, and E. R. Hancock, \href{https://academic.oup.com/comnet/article-abstract/5/6/858/3925042?redirectedFrom=fulltext}{J. Complex Netw. \textbf{5}, 858 (2017).}
\bibitem{braun1} S. L. Braunstein, S. Ghosh, and S. Severini, \href{https://link.springer.com/article/10.1007/s00026-006-0289-3}{Ann. Comb. \textbf{10}, 291 (2006).}
\bibitem{paserini1} F. Passerini and S. Severini, \href{http://dx.doi.org/10.4018/jats.2009071005}{Int. J. Agent Technol. Syst. \textbf{1}, 58 (2009).}
\bibitem{anand1} K. Anand, G. Bianconi, and S. Severini, \href{http://dx.doi.org/10.1103/PhysRevE.83.036109}{Phys. Rev. E \textbf{83}, 036109 (2011).}
\bibitem{minello4} G. Minello, L. Rossi, and A. Torsello, \href{https://academic.oup.com/comnet/article-abstract/7/4/491/5208410?redirectedFrom=fulltext}{J. Complex Netw. \textbf{7}, 491 (2019).}
\bibitem{gabrielli1} A. Gabrielli, R. Mastrandrea, G. Caldarelli, and G. Cimini, \href{http://dx.doi.org/10.1103/PhysRevE.99.030301}{Phys. Rev. E \textbf{99}, 030301 (2019).}
\bibitem{metz1} C. Metzig and C. Colijin, \href{https://www.mdpi.com/1099-4300/22/3/312}{Entropy \textbf{22}, 312 (2020).}
\bibitem{bianconi3} G. Bianconi, \href{http://dx.doi.org/10.1103/PhysRevE.66.036116}{Phys. Rev. E \textbf{66}, 036116 (2002).}
\bibitem{shen1} Y. Shen, D. L. Zhu, and W. M. Liu, \href{http://dx.doi.org/10.1088/0256-307X/22/5/072/meta}{Chinese Physics Letters \textbf{22}, 1281 (2005).}
\bibitem{moura1} A. P. S. de Moura, \href{http://dx.doi.org/10.1103/PhysRevE.71.066114}{Phys. Rev. E \textbf{71}, 066114 (2005).}
\bibitem{murphy1} C. Murphy, A. Allard, E. Laurence, G. St-Onge, and L. J. Dubé, \href{http://dx.doi.org/10.1103/PhysRevE.97.032309}{Phys. Rev. E \textbf{97}, 032309 (2018).}
 
\bibitem{svaiter1} B. F. Svaiter and N. F. Svaiter, \href{https://doi.org/10.1142/S0217751X1650144X}{Int. J. Mod. Phys. A \textbf{31}, 1650144 (2016)}.   
\bibitem{mezard1} M. Mézard, G. Parisi, and M. A. Virasoro, \href{https://iopscience.iop.org/article/10.1209/0295-5075/1/2/006/meta}{EPL \textbf{1}, 2 (1986)}.
\bibitem{mezard2} M. Mézard and A. Montanari, \emph{Information, Physics, and Computation} (Oxford University Press, 2009).
\bibitem{edw1} S. F. Edwards and P. W. Anderson, \href{https://iopscience.iop.org/article/10.1088/0305-4608/5/5/017/meta}{J. Phys. F \textbf{5}, 017 (1975)}. 
\bibitem{parisi1} G. Parisi, \href{https://doi.org/10.1103/PhysRevLett.43.1754}{Phys. Rev. Lett. \textbf{43}, 1754 (1979)}.
\bibitem{parisi2} G. Parisi, \href{https://iopscience.iop.org/article/10.1088/0305-4470/13/4/009}{J. Phys. A \textbf{13}, 009 (1980)}.
\bibitem{parisi3} G. Parisi, \href{https://iopscience.iop.org/article/10.1088/0305-4470/13/5/047}{J. Phys. A \textbf{13}, 047 (1980)}.
\bibitem{parisi4} G. Parisi, \href{https://doi.org/10.1103/PhysRevLett.50.1946}{Phys. Rev. Lett. \textbf{50}, 1946 (1983)}.
\bibitem{svaiter2} R. A. Diaz, G. Menezes, N. F. Svaiter, and C. A. D. Zarro, \href{https://doi.org/10.1103/PhysRevD.96.065012}{Phys. Rev. D \textbf{96}, 065012 (2017)}. 
\bibitem{svaiter3} R. A. Diaz, N. F. Svaiter, G. Krein, and C. A. D. Zarro, \href{https://doi.org/10.1103/PhysRevD.97.065017}{Phys. Rev. D \textbf{97}, 065017 (2018)}. 
\bibitem{svaiter4} R. A. Diaz, G. Krein, A. Saldivar, N. F. Svaiter, and C. A. D. Zarro, \href{https://iopscience.iop.org/article/10.1088/1751-8121/ab4687/meta}{J. Phys. A \textbf{52}, 445401 (2019)}.
\bibitem{svaiter5} M. S. Soares, N. F. Svaiter, and C. A. D. Zarro, \href{https://iopscience.iop.org/article/10.1088/1361-6382/ab4fd3}{Classical Quantum Gravity \textbf{37}, 065024 (2020)}.   
\bibitem{svaiter6} R. A. Diaz, C. D. Rodríguez-Camargo, and N. F. Svaiter, \href{https://doi.org/10.3390/polym12051066}{Polymers \textbf{12}, 1066 (2020)}.   
\bibitem{svaiter7} G. Menezes, N. F. Svaiter, and C. A. D. Zarro, \href{https://doi.org/10.1103/PhysRevA.96.062119}{Phys. Rev. A \textbf{96}, 062119 (2017)}.  
\bibitem{estrada2} E. Estrada and N. Hatano, \href{https://doi.org/10.1103/PhysRevE.77.036111}{Phys. Rev. E \textbf{77}, 036111 (2008)}.  
\bibitem{marshall1} A. W. Marshall, I. Olkin, and B. C. Arnold, \emph{Inequalities: Theory of Majorization and Its Applications} (Springer, New York, 2011).
\bibitem{chang11} C. Yi-Fang, Galilean Electrodynamics \textbf{21}, 112 (2010).
\bibitem{riemann1} B. Riemann, \href{https://www.claymath.org/sites/default/files/zeta.pdf}{Monatsb. der Berliner Akad. \textbf{1858/60}, 671}.
\bibitem{ing1} A. E. Ingham, \emph{The Distribution of Prime Numbers} (Cambridge University Press, UK, 1990).
\bibitem{ulandau1} E. Landau and A. Walfisz, \href{https://link.springer.com/article/10.1007/BF03014596}{Rend. Circ. Mat. di Palermo \textbf{44}, 82 (1920)}.
\bibitem{froberg1} C.-E. Fröberg, \href{https://link.springer.com/article/10.1007/BF01933420}{BIT \textbf{8}, 187 (1968)}.
\bibitem{blau1} S. K. Blau, M. Visser, and A. Wipf, \href{https://doi.org/10.1016/0550-3213(88)90059-4}{Nuclear Phys. B \textbf{310}, 163 (1988)}.
\bibitem{haw1} S. W. Hawking, \href{https://link.springer.com/article/10.1007/BF01626516}{Commun. Math. Phys. \textbf{55}, 133 (1977)}.
\bibitem{hajli1} M. Hajli, \href{https://doi.org/10.1016/j.jnt.2019.07.020}{J. Number Theory \textbf{208}, 120 (2020)}.
\bibitem{ramsey1} F. P. Ramsey, \href{http://www.cs.umd.edu/~gasarch/BLOGPAPERS/ramseyorig.pdf}{Proc. Lond. Math. Soc. \textbf{s2-30}, 264 (1930)}. 
\bibitem{lamaison1} A. Lamaison, \href{http://www.iam.fmph.uniba.sk/amuc/ojs/index.php/amuc/article/view/1194/742}{Acta Math. Univ. Comenian. (N.S.) \textbf{88}, 897 (2019)}.  
\bibitem{causey1} R. M. Causey and C. Doebele, \href{https://www.impan.pl/en/publishing-house/journals-and-series/fundamenta-mathematicae/all/248/2/113023/a-ramsey-theorem-for-pairs-in-trees}{Fund. Math. \textbf{248}, 147 (2020)}.
\bibitem{choi1} I. Choi, M. Furuya, R. Kim, and B. Park, \href{https://doi.org/10.1016/j.disc.2019.111648}{Discrete Math. \textbf{343}, 111648 (2020)}.
\bibitem{charles1} Z. B. Charles, M. Farber, C. R. Johnson, and K.-S. Lee, \href{https://doi.org/10.1137/130904624}{SIAM J. Matrix Anal. Appl. \textbf{34}, 1384 (2013)}.
\bibitem{farber1} M. Farber and C. R. Johnson, \href{https://doi.org/10.1080/03081087.2013.872242}{Linear Multilinear Algebra \textbf{63}, 423 (2015)}.
\bibitem{johnson1} C. R. Johnson and R. B. Reams, \href{https://doi.org/10.1515/spma-2016-0007}{Spec. Matrices \textbf{4}, 67 (2016)}. 
\bibitem{streit1} A. Streitwieser, in \href{https://pubs.acs.org/doi/abs/10.1021/bk-2013-1122.ch009}{\emph{Pioneers of Quantum Chemistry}, edited by E. T. Strom and A. K. Wilson (American Chemical Society, 2013)}.
\bibitem{coulson1} C. A. Coulson, B. O’Leary, and R. B. Mallion, \emph{Hückel Theory for Organic Chemists} (Academic, New York, 1978).
\bibitem{estrada23} E. Estrada, \href{https://iopscience.iop.org/article/10.1209/epl/i2005-10441-3/meta}{Europhys. Lett. \textbf{73}, 649 (2006).}
\bibitem{estrada24} E. Estrada, \href{https://doi.org/10.1016/j.jtbi.2006.08.002}{J. Theor. Biol. \textbf{244}, 296 (2007).}
\bibitem{estrada25} G. Palla, I. Derényi, I. Farkas, and T. Vicsek, \href{https://www.nature.com/articles/nature03607}{Nature \textbf{435}, 814 (2005).}
\bibitem{ma1} S.-K. Ma and J. Rudnick, \href{https://doi.org/10.1103/PhysRevLett.40.589}{Phys. Rev. Lett. \textbf{40}, 589 (1978)}. 
\bibitem{sherr1} D. Sherrington, \href{https://doi.org/10.1103/PhysRevB.22.5553}{Phys. Rev. B \textbf{22}, 5553 (1980)}.
\bibitem{tarjus1} G. Tarjus and V. Dotsenko, \href{https://iopscience.iop.org/article/10.1088/0305-4470/35/7/311/meta}{J. Phys. A \textbf{35}, 1627 (2002)}.
\bibitem{piit1} L. P. Pitaevskii, Sov. Phys.-JETP \textbf{13}, 451 (1961).
\bibitem{gross11} E. P. Gross, \href{https://doi.org/10.1063/1.1703944}{J. Math. Phys. \textbf{4}, 195 (1963)}.
\bibitem{calzeta0} E. Calzetta, B. L. Hu, and E. Verdaguer, \href{https://doi.org/10.1142/S0217979207045475}{Int. J. Modern Phys. B \textbf{21}, 4239 (2007)}.
\bibitem{syman1} K. Symanzik, \href{https://doi.org/10.1063/1.1704960}{J. Math. Phys. \textbf{7}, 510 (1966)}.
\bibitem{hartle1} J. B. Hartle and S. W. Hawking, \href{https://doi.org/10.1103/PhysRevD.13.2188}{Phys. Rev. D \textbf{13}, 2188 (1976)}.
\bibitem{gibbons1} G. W. Gibbons and S. W. Hawking, \href{https://doi.org/10.1103/PhysRevD.15.2752}{Phys. Rev. D \textbf{15}, 2752 (1977)}.
\bibitem{calzeta1} E. Calzetta and B. L. Hu, \href{https://doi.org/10.1103/PhysRevD.49.6636}{Phys. Rev. D \textbf{49}, 6636 (1994)}.
\bibitem{calzeta2} E. Calzetta, \href{https://iopscience.iop.org/article/10.1088/0264-9381/29/14/143001/meta}{Classical Quantum Gravity \textbf{29}, 143001 (2012)}.
\bibitem{biamonte1} J. Biamonte, M. Faccin, and M. De Domenico, \href{https://www.nature.com/articles/s42005-019-0152-6}{Nat. Commun. Phys. \textbf{2}, 53 (2019)}.
\bibitem{elias1} A. D. Verga and R. G. Elías, \href{https://doi.org/10.1103/PhysRevE.100.062137}{Phys. Rev. E \textbf{100}, 062137 (2019)}.
\bibitem{potes1} A. N. Poteshman, E. Tang, L. Papadopoulos, D. S. Bassett, and L. C. Bassett, \href{https://iopscience.iop.org/article/10.1088/1367-2630/ab5c9f/meta}{N. J. Phys. \textbf{21}, 123049 (2019)}.
\bibitem{garnerone1} S. Garnerone, P. Giorda, and P. Zanardi, \href{https://iopscience.iop.org/article/10.1088/1367-2630/14/1/013011/meta}{N. J. Phys. \textbf{14}, 013011 (2012)}.
\bibitem{perseguers1} S. Perseguers, G. J. Lapeyre, D. Cavalcanti, M. Lewenstein, and A. Acín, \href{https://iopscience.iop.org/article/10.1088/0034-4885/76/9/096001/meta}{Rep. Prog. Phys. \textbf{76}, 096001 (2013)}.
\bibitem{kolli1} A. Kolli, E. J. O'Reilly, G. D. Scholes, and A. Olaya-Castro, \href{https://doi.org/10.1063/1.4764100}{J. Chem. Phys. \textbf{137}, 174109 (2012)}.
\bibitem{stones1} R. Stones and A. Olaya-Castro, \href{https://doi.org/10.1016/j.chempr.2016.11.014}{Chem \textbf{1}, 822 (2016)}.
\bibitem{curut1} C. Curutchet and B. Mennucci, \href{https://pubs.acs.org/doi/abs/10.1021/acs.chemrev.5b00700}{Chem. Rev. \textbf{117}, 294 (2017)}.
\bibitem{qin1} M. Qin, H. Z. Shen, X. L. Zhao, and X. X. Yi, \href{https://doi.org/10.1103/PhysRevA.96.012125}{Phys. Rev. A \textbf{96}, 012125 (2017)}.
\bibitem{plenio1} A. Streltsov, G. Adesso, and M. B. Plenio, \href{https://doi.org/10.1103/RevModPhys.89.041003}{Rev. Mod. Phys. \textbf{89}, 041003 (2017)}.
\bibitem{eduardom1} C. Caro-Ruiz, J. Ma, D. J. Hill, A. Pavas, and E. Mojica-Nava, \href{https://doi.org/10.1016/j.segan.2020.100302}{Sustain. Energy Grids \textbf{21}, 100302 (2020).}
\end{thebibliography}
\end{document}